\documentclass[aps,prd,onecolumn,groupedaddress,showpacs,nofootinbib,amssymb]{revtex4-2}
\usepackage[dvips]{graphicx}
\usepackage{amssymb}
\usepackage{amsmath}
\usepackage{hyperref}
\usepackage{graphicx,color}
\usepackage{amsfonts}
\usepackage{bm}
\usepackage{cancel}
\usepackage{comment}

\newcommand\be{\begin{equation}}
\newcommand\ee{\end{equation}}

\allowdisplaybreaks[4]

\begin{document}

\tolerance=5000

\title{Inflationary Attractors Predictions for Static Neutron Stars in the Mass-Gap Region}
\author{S.D. Odintsov,$^{1,2}$}
\author{V.K. Oikonomou,$^{3}$}
\email{odintsov@ice.cat}
\email{voikonomou@gapps.auth.gr,v.k.oikonomou1979@gmail.com}
\affiliation{$^{1)}$ICREA, Passeig Luis Companys, 23, 08010
Barcelona, Spain \\
$^{2)}$ Institute of Space Sciences (ICE, CSIC) C. Can Magrans
s/n, 08193 Barcelona, Spain
\\
$^{3)}$Department of Physics, Aristotle University of
Thessaloniki, Thessaloniki 54124, Greece\\}

\tolerance=5000

\begin{abstract}
In this work we study static neutron stars in the context of
several inflationary models which are popular in cosmology. These
inflationary models are non-minimally coupled scalar theories
which yield a viable inflationary phenomenology in both Jordan and
Einstein frames. By considering the constraints from inflationary
theories, which basically determine the values of the potential
strength, usually considered as a free parameter in astrophysical
neutron star works, we construct and solve the
Tolman-Oppenheimer-Volkoff equations using a solid python-3 LSODA
integrator. For our study we consider several popular inflationary
models, such as the universal attractors, the $R^p$ attractors
(three distinct model values), the induced inflation, the
quadratic inflation, the Higgs inflation and the $a$-attractors
(two distinct model values) and for the following popular
equations of state the WFF1, the SLy, the APR, the MS1, the AP3,
the AP4, the ENG, the MPA1 and the MS1b. We construct the $M-R$
diagram and we confront the resulting theory with theoretical and
observational constraints. As we demonstrate, remarkably, all the
neutron stars produced by all the inflationary models we
considered are compatible with all the constraints for the MPA1
equation of state. It is notable that for this particular equation
of state, the maximum masses of the neutron stars are in the
mass-gap region with $M>2.5M_{\odot}$, but lower than the 3 solar
masses causal limit. Another important feature of our work is that
it may be possible to discriminate inflationary attractors which
at the cosmological level are indistinguishable using the $M-R$
graphs of static neutron stars, however we point out the
limitations in discriminating the inflationary attractors. Also we
show that the WFF1, MS1 and MS1b seem to be entirely ruled out,
regarding a viable description of static neutron stars. We also
make the observation that as the NICER constraints are pushed
towards larger radii, as for example in the case of the black
widow pulsar PSR J0952-0607, it seems that equations of state that
produce neutron stars with maximum masses in the mass gap region,
with $M>2.5M_{\odot}$, but lower than the 3 solar masses causal
limit, are favored and are compatible with the modified NICER
constraints. Finally we question the ability of the MPA1 equation
of state to pass all the theoretical and observational constraints
and we impose the question whether this equation of state plays
any fundamental role in static neutron star physics.
\end{abstract}


\maketitle

\section*{Introduction}

\begin{figure}[h!]
\centering
\includegraphics[width=30pc]{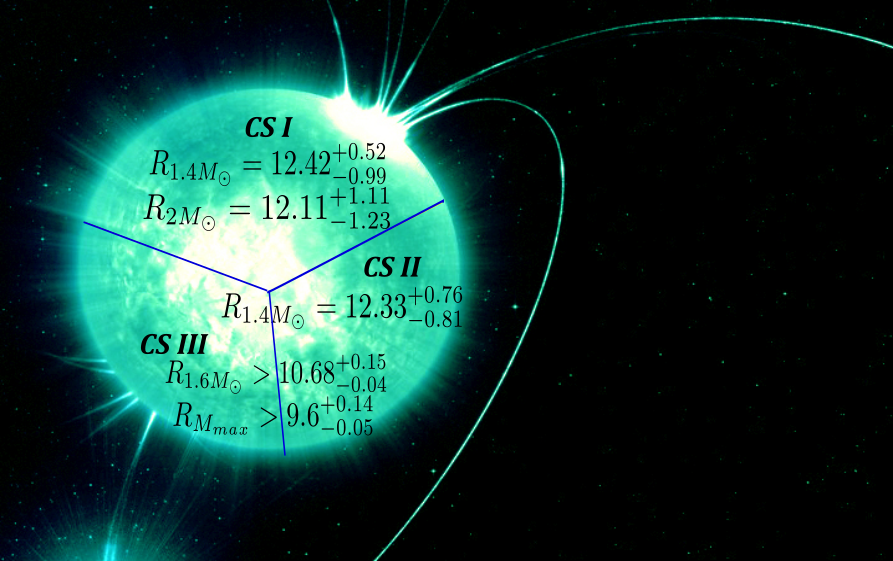}
\caption{An illustrative presentation of the constraints CSI
\cite{Altiparmak:2022bke} $R_{1.4M_{\odot}}=12.42^{+0.52}_{-0.99}$
and $R_{2M_{\odot}}=12.11^{+1.11}_{-1.23}\,$km, the constraint
CSII \cite{Raaijmakers:2021uju} in which case
$R_{1.4M_{\odot}}=12.33^{+0.76}_{-0.81}\,\mathrm{km}$ and the
constraint CSIII \cite{Bauswein:2017vtn} in which case the radius
of an $1.6M_{\odot}$ mass NS must be larger than
$R_{1.6M_{\odot}}>10.68^{+0.15}_{-0.04}\,$km and for the maximum
mass of a NS, the corresponding radius must be large than
$R_{M_{max}}>9.6^{+0.14}_{-0.03}\,$km. This illustrative figure is
edited based on a public image of ESO, which can be found freely
in Credit: ESO/L.Cal\c{c}ada:
\url{https://www.eso.org/public/images/eso0831a/.}} \label{plotcs}
\end{figure}

Direct gravitational waves observations have utterly changed the
perspective of physicists on how they understand the Universe.
Starting with the kilonova event GW170817
\cite{TheLIGOScientific:2017qsa,Abbott:2020khf}, theorists coming
from cosmology and particle physics had to revise which theories
can describe a realistic theory of cosmology, narrowing down
significantly the available theories, since the event excluded all
massive gravity theories
\cite{Ezquiaga:2017ekz,Baker:2017hug,Creminelli:2017sry,Sakstein:2017xjx}.
Refinements to theories that predict a massive tensor spectrum
were offered in the literature
\cite{Odintsov:2020sqy,Oikonomou:2021kql,Oikonomou:2022ksx},
however, everything in the literature now is abundantly clear
which theories are cosmologically viable. Thus a single kilonova
event already had a serious impact on the phenomenology of
cosmological theories. It is highly anticipated from future
observations of Neutron Stars (NSs) to see whether alternative
information can be gained from gravitational systems in extreme
gravitational environments. The main question is whether modified
gravity \cite{reviews1,reviews2,reviews3,reviews4} can play some
fundamental role in extreme gravity environments, such as NSs and
cosmology. At the moment, we only have hints and indications that
this might be the case, but these hints are premature to justify a
concrete definite answer to the problem. More observations are
required on these issues. To date, there are two main events and
observations that point in the direction of having heavy NSs, the
GW190814 event \cite{LIGOScientific:2020zkf} which points having
NSs inside the mass-gap region $M>2.5M_{\odot}$, and also the
black-widow binary pulsar PSR J0952-0607 with mass $M=2.35\pm
0.17$ \cite{Romani:2022jhd}. The latter does not predict a neutron
star (NS) mass inside the mass-gap region, but still the NS is
quite heavier that the NICER predictions of $\sim 2$ solar masses.
These are the hints that we believe nature might have some near
future surprises with NSs inside the mass-gap region, for which
the simple general relativistic (GR) description might not
suffice. So for heavy NS the only optimal and Occam's razor based
approach is to assume that modified gravity in some form of it
controls the hydrodynamic stability of the NS and drives its
maximum mass to regions which are extremely unreachable by GR,
even for the stiffest and not justifiable equations of state
(EoSs). From these considerations, it is apparent that NSs
\cite{Haensel:2007yy,Friedman:2013xza,Baym:2017whm,Lattimer:2004pg,Olmo:2019flu})
and similar astrophysical compact objects have become the test bed
of future gravitational and to some extent particle physics
theories. These are a virtual laboratory at the sky in which high
energy particle physics, particle astrophysics and cosmology
theories can be tested. Obviously we live in the golden era of NSs
and future illuminating observations are highly anticipated. Many
distinct scientific areas can be studied in NSs, for example
nuclear theories of extreme matter conditions
\cite{Lattimer:2012nd,Steiner:2011ft,Horowitz:2005zb,Watanabe:2000rj,Shen:1998gq,Xu:2009vi,Hebeler:2013nza,Mendoza-Temis:2014mja,Ho:2014pta,Kanakis-Pegios:2020kzp,Tsaloukidis:2022rus},
high energy theoretical and particle physics
\cite{Buschmann:2019pfp,Safdi:2018oeu,Hook:2018iia,Edwards:2020afl,Nurmi:2021xds},
studies in modified gravity of various forms,
\cite{Astashenok:2020qds,Astashenok:2021peo,Capozziello:2015yza,Astashenok:2014nua,Astashenok:2014pua,Astashenok:2013vza,Arapoglu:2010rz,Panotopoulos:2021sbf,Lobato:2020fxt,Numajiri:2021nsc}
and finally theoretical astrophysics studies,
\cite{Altiparmak:2022bke,Bauswein:2020kor,Vretinaris:2019spn,Bauswein:2020aag,Bauswein:2017vtn,Most:2018hfd,Rezzolla:2017aly,Nathanail:2021tay,Koppel:2019pys,Raaijmakers:2021uju,Most:2020exl,Ecker:2022dlg,Jiang:2022tps}.
Modified gravity can eventually play a fundamental role in NS
physics, but this is still questionable. Indeed, in cosmology one
of the alternative and theoretically consistent way to describe
dark energy in all the experimentally allowed values of the dark
energy EoS, can be done by modified gravity in its various forms,
while in the context of simple GR, one should resort to phantom
scalar fields, a rather unappealing description of nature, for the
moment at least. Apart from that, if the inflationary era ever
existed, the GR description of inflation with a scalar field has
many shortcomings, the most important being the excess of the
inflaton's couplings to the Standard Model particles, which are
needed for the thermalization of the Universe. In modified
gravity, the instability of the inflationary attractor at the end
of inflation, causes oscillatory solutions to the curvature that
may directly reheat the particle content of the Universe directly,
without resorting to thermalization via particle interactions.
Thus modified gravity in its various forms, seems a viable
description of nature, at least for the time being. In the
literature there exists a vast stream of articles studying NSs in
the context of modified gravity
\cite{Astashenok:2014nua,Astashenok:2014pua} and also in the
context of scalar-tensor theories
\cite{Pani:2014jra,Staykov:2014mwa,Horbatsch:2015bua,Silva:2014fca,Doneva:2013qva,Xu:2020vbs,Salgado:1998sg,Shibata:2013pra,Arapoglu:2019mun,Ramazanoglu:2016kul,AltahaMotahar:2019ekm,Chew:2019lsa,Blazquez-Salcedo:2020ibb,Motahar:2017blm,Odintsov:2021qbq,Odintsov:2021nqa,Oikonomou:2021iid,Pretel:2022rwx,Pretel:2022plg,Cuzinatto:2016ehv,Oikonomou:2023dgu}.
Among the many scalar-tensor theories, the inflationary theories
are deemed the most important from multiple aspects, mainly though
because they can generate a viable inflationary era compatible
with the Planck data \cite{Akrami:2018odb}. Here we shall consider
an important class of inflationary theories, the inflationary
attractors
\cite{alpha0,alpha1,alpha2,alpha3,alpha4,alpha5,alpha6,alpha7,alpha7a,alpha8,alpha9,alpha10,alpha11,alpha12,alpha13,alpha14,alpha15,alpha16,alpha17,alpha18,alpha19,alpha20,alpha21,alpha22,alpha23,alpha24,alpha25,alpha26,alpha27,alpha28,alpha29,alpha30,alpha31,alpha32,alpha33,alpha34,alpha35,alpha36,alpha37}.
These theories are called attractors because although they
originate for a different Jordan frame scalar field theory, they
result to the same inflationary phenomenology in the Einstein
frame, so essentially they are attracted to the same phenomenology
in the Einstein frame. By taking also into account the constraints
from inflationary theories, which essentially  determine the
values of the potential strength, the coefficient of the scalar
potential, usually considered as a free parameter in astrophysical
NS works, we construct and solve the Tolman-Oppenheimer-Volkoff
(TOV) equations using a well known python-3 LSODA integrator. For
our study we shall consider several popular and viable
inflationary models, such as the universal attractors, the $R^p$
attractors (three distinct model values), the induced inflation,
the quadratic inflation, the Higgs inflation and the
$a$-attractors (two distinct model values) and regarding the EoS,
we shall use nine different EoSs in order to describe the nuclear
matter inside the NS. We shall use a piecewise polytropic approach
\cite{Read:2008iy,Read:2009yp} for all the EoS we will use, in
which the low-density part can be one of the following EoSs: the
SLy \cite{Douchin:2001sv} which is a potential method EoS, the
AP3-AP4 \cite{Akmal:1998cf} which is a variational method EoS, the
WFF1 \cite{Wiringa:1988tp} which is also a variational method EoS,
the ENG \cite{Engvik:1995gn} and the MPA1 \cite{Muther:1987xaa}
which are relativistic EoSs, the MS1 and MS1b
\cite{Mueller:1996pm} which are relativistic mean field theory
EoSs, with the MS1b being identical to the MS1 with a low symmetry
energy of 25$\,$MeV and finally the APR EoS \cite{Akmal:1997ft}.
Our aim for solving the TOV equations is to obtain the masses and
radii of NSs in the Jordan frame eventually, and construct the
$M-R$ graphs. Regarding the gravitational mass, we shall consider
the Arnowitt-Deser-Misner (ADM) masses \cite{Arnowitt:1960zzc} and
we shall provide an explicit formula on this. We shall eventually
confront our results with the observational data, using
theoretical and observational constraints. Particularly, we shall
consider three types of constraints, which we call CSI, CSII and
CSIII.  The CSI was firstly introduced in Ref.
\cite{Altiparmak:2022bke} and indicates that the radius of an
$1.4M_{\odot}$ mass NS must be
$R_{1.4M_{\odot}}=12.42^{+0.52}_{-0.99}$ while the radius of an
$2M_{\odot}$ mass NS must be
$R_{2M_{\odot}}=12.11^{+1.11}_{-1.23}\,$km. The second constraint
we shall consider is CSII and was introduced in Ref.
\cite{Raaijmakers:2021uju} and indicates that the radius of an
$1.4M_{\odot}$ mass NS must be
$R_{1.4M_{\odot}}=12.33^{+0.76}_{-0.81}\,\mathrm{km}$.
Furthermore, we shall consider a third constraint, namely CSIII,
which was firstly introduced in Ref. \cite{Bauswein:2017vtn} and
indicates that the radius of an $1.6M_{\odot}$ mass NS must be
larger than $R_{1.6M_{\odot}}>10.68^{+0.15}_{-0.04}\,$km, while
when the radius of a NS with maximum mass is considered, it must
be larger than $R_{M_{max}}>9.6^{+0.14}_{-0.03}\,$km. A graphical
representation of the constraints CSI, CSII and CSIII can be found
in Fig. \ref{plotcs}.

In addition to these constraints, we shall also include the NICER
I constraints \cite{Miller:2021qha} valid for $M=1.4M_{\odot}$
NSs, which constraints the radius to be
$R_{1.4M_{\odot}}=11.34-13.23\,$km. Furthermore, we shall also
consider a theoretical refinement of the NICER constraint recently
introduced in \cite{Ecker:2022dlg} by taking into account the
heavy black-widow binary pulsar PSR J0952-0607 with mass
$M=2.35\pm 0.17$ \cite{Romani:2022jhd}. We shall call this
constraint NICER II. The NICER II constraints the radius of a
$M=1.4M_{\odot}$ to be $R_{1.4M_{\odot}}=12.33-13.25\,$km. After
performing a thorough analysis of all the models and EoS, we show
that the MPA1 EoS, shows remarkable compatibility properties with
all the constraints. We discuss in detail the phenomenological
implications of our results.

\section{Neutron Stars Physics and Cosmological Aspects of Inflationary Attractors: Notation, Formalism and EoSs}

In general, there is a different notation between neutron star
physics applications of non-minimally coupled scalar theories and
inflationary non-minimally coupled theories. In this section we
shall bridge the gap between the two distinct approaches and we
shall make a direct correspondence between the notation of the two
approaches. We first consider the inflationary attractors context,
see
\cite{alpha0,alpha1,alpha2,alpha3,alpha4,alpha5,alpha6,alpha7,alpha7a,alpha8,alpha9,alpha10,alpha11,alpha12,alpha13,alpha14,alpha15,alpha16,alpha17,alpha18,alpha19,alpha20,alpha21,alpha22,alpha23,alpha24,alpha25,alpha26,alpha27,alpha28,alpha29,alpha30,alpha31,alpha32,alpha33,alpha34,alpha35,alpha36,alpha37}.
See also Refs.
\cite{Kaiser:1994vs,valerio,Faraoni:2013igs,Buck:2010sv} regarding
the conformal transformations in inflationary theories. We first
consider the Jordan frame gravitational action of a non-minimally
coupled scalar field,
\begin{equation}\label{c1intro}
\mathcal{S}_J=\int
d^4x\Big{[}f(\phi)R-\frac{\omega(\phi)}{2}g^{\mu
\nu}\partial_{\mu}\phi\partial_{\nu}\phi-U(\phi)\Big{]}+S_m(g_{\mu
\nu},\psi_m)\, ,
\end{equation}
where we also took into account the presence of perfect matter
fluids quantified by the action $S_m(g_{\mu \nu},\psi_m)$, and we
denote the pressure of the perfect matter fluids as $P$ while
their energy density as $\epsilon$. For cosmological applications
it is customary to use natural units ($c=\hbar=1$), while in
contrast for NS applications one uses Geometrized units. In the
minimally coupled version of the Jordan frame case, one has the
Einstein frame theory at hand, with
\begin{equation}\label{c2intro}
f(\phi)=\frac{1}{16 \pi G}=\frac{M_p^2}{2}\, ,
\end{equation}
and the reduced Planck mass in natural units is defined as,
\begin{equation}\label{c3intro}
M_p=\frac{1}{\sqrt{8\pi G}}\, ,
\end{equation}
with $G$ being the Jordan frame Newton's gravitational constant.
Regarding the non-minimally coupled Jordan frame theory, we
perform the conformal transformation,
\begin{equation}\label{c4intro}
\tilde{g}_{\mu \nu}=\Omega^2g_{\mu \nu}\, ,
\end{equation}
and one obtains the Einstein frame gravitational action, and we
use the ``tilde'' to denote the Einstein frame physical
quantities. In the Einstein frame, one obtains the minimally
coupled scalar field theory by choosing the function $\Omega$
entering the conformal transformation as follows
\cite{Kaiser:1994vs,valerio},
\begin{equation}\label{c6intro}
\Omega^2=\frac{2}{M_p^2}f(\phi)\, ,
\end{equation}
therefore, by making the conformal transformation in the action
(\ref{c1intro}), and by also making the choice (\ref{c6intro}),
the Einstein frame action is obtained, which reads,
\begin{equation}\label{c12intro}
\mathcal{S}_E=\int
d^4x\sqrt{-\tilde{g}}\Big{[}\frac{M_p^2}{2}\tilde{R}-\frac{\zeta
(\phi)}{2} \tilde{g}^{\mu \nu }\tilde{\partial}_{\mu}\phi
\tilde{\partial}_{\nu}\phi-V(\phi)\Big{]}+S_m(\Omega^{-2}\tilde{g}_{\mu
\nu},\psi_m)\, ,
\end{equation}
and the Einstein frame scalar field potential $V(\phi)$ is related
the corresponding Jordan frame scalar field potential $U(\phi)$ in
the following way,
\begin{equation}\label{c13intro}
V(\phi)=\frac{U(\phi)}{\Omega^4}\, ,
\end{equation}
and furthermore the definition of the function $\zeta(\phi)$ is,
\begin{equation}\label{c14intro}
\zeta
(\phi)=\frac{M_p^2}{2}\Big{(}\frac{3\Big{(}\frac{df}{d\phi}\Big{)}^2}{f^2}+\frac{2\omega(\phi)}{f}\Big{)}\,
.
\end{equation}
By appropriately choosing the kinetic term of the scalar field,
namely the function $\zeta(\phi)$, one can render the Einstein
frame scalar theory canonical. This can be done in terms of the
following transformation,
\begin{equation}\label{c15intro}
\Big{(}\frac{d\varphi}{d \phi}\Big{)} =\sqrt{\zeta(\phi)}\, ,
\end{equation}
and thus, the Einstein frame scalar field theory acquires its
well-known canonical form,
\begin{equation}\label{c17intro}
\mathcal{S}_E=\int
d^4x\sqrt{-\tilde{g}}\Big{[}\frac{M_p^2}{2}\tilde{R}-\frac{1}{2}\tilde{g}^{\mu
\nu } \tilde{\partial}_{\mu}\varphi
\tilde{\partial}_{\nu}\varphi-V(\varphi)\Big{]}+S_m(\Omega^2\tilde{g}_{\mu
\nu},\psi_m)
\end{equation}
with,
\begin{equation}\label{c18intro}
V(\varphi)=\frac{U(\varphi)}{\Omega^4}=\frac{U(\varphi)}{4
M_p^4f^2}\, .
\end{equation}
An important feature of the Einstein frame theory is the fact that
the matter fluids are no longer perfect fluids, and fact that can
be seen by the energy momentum tensor which satisfies,
\begin{equation}\label{c24intro}
\tilde{\partial}^{\mu}\tilde{T}_{\mu \nu}=-\frac{d}{d\varphi}[\ln
\Omega]\tilde{T}\tilde{\partial}_{\nu}\phi\, .
\end{equation}
Furthermore, the energy density and the pressure of the matter
fluids have the following transformation properties between the
Jordan and the Einstein frame,
\begin{equation}\label{c28intro}
\tilde{\varepsilon}=\Omega^{-4}(\varphi)\varepsilon,\,\,\,\tilde{P}=\Omega^{-4}(\varphi)P\,
,
\end{equation}
where recall that the ``tilde'' denotes Einstein frame physical
quantities. A useful notation for the Einstein frame scalar field
theory is the following,
\begin{equation}\label{einsteinframeactionintro}
\mathcal{S}_E=\int
d^4x\sqrt{-\tilde{g}}\Big{[}\frac{M_p^2}{2}\tilde{R}-\frac{1}{2}\tilde{g}^{\mu
\nu } \tilde{\partial}_{\mu}\varphi
\tilde{\partial}_{\nu}\varphi-V(\varphi)\Big{]}\, ,
\end{equation}
which can be further cast as follows,
\begin{equation}\label{einsteinframeactioninflationnsintro}
\mathcal{S}_E=\int d^4x\sqrt{-\tilde{g}}\Big{[}\frac{1}{16\pi
G}\tilde{R}-\frac{1}{2}\tilde{g}^{\mu \nu }
\tilde{\partial}_{\mu}\varphi
\tilde{\partial}_{\nu}\varphi-\frac{16\pi G V(\varphi)}{16\pi
G}\Big{]}\, ,
\end{equation}
with $M_p^2=\frac{1}{8\pi G}$. The expression of
(\ref{einsteinframeactioninflationnsintro}) is useful for NS
physics. We need to note that all these theories seen as Jordan
frame theories generally violate the weak equivalence principle,
see for example \cite{ref1,ref2}. Specifically, the Jordan frame
scalar theory transformed in the Einstein frame generally exactly
violates the weak equivalence principle \cite{ref1} unless the
coupling of the scalar field to gravity in the Jordan frame is
conformal. Let us note that if the scalar field disappears from
the Universe due to its decays during the radiation domination era
to the Standard model particles, which is actually the mechanism
of reheating in inflaton based inflationary theories, the
violation of the equivalence principle has eventually disappeared
from the Universe and has no effect on the Solar System
experiments. Else, a fifth force should be present so in these
theories the post-Newtonian parameters are affected, but to date
for most scalar-tensor theories no important  violation of the
equivalence principle is predicted. Such strong deviations would
be present for theories in which $\frac{d^2 \ln A(\varphi)}{d
\varphi^2}\leq -4$ (see below Eq.
(\ref{alphaofvarphigeneraldefintro}) for a definition), a
constraint which does not hold true for the theories which we
shall study. Caution is needed for theories in which the coupling
function $A(\varphi)$ is sinusoidal. Of course all these issues
could be avoided if the inflationary theory is generated
geometrically via some $f(R)$ gravity, but still we should have
these issues into account if one uses the conformally transformed
version of the theory in the Einstein frame.

Now let us proceed to the formalism and notation of non-minimally
coupled scalar field theories in the Jordan frame, customary for
NSs applications. We shall adopt the notation of
\cite{Pani:2014jra} and we shall derive the TOV equations in
general format for scalar field theories with potential. In
theoretical astrophysics contexts it is customary to use
Geometrized units ($G=c=1$), so let us express the Jordan frame
gravitational action of a non-minimally coupled inflationary
theory we presented previously, in an theoretical astrophysics
context, which is the following,
\begin{equation}\label{taintro}
\mathcal{S}=\int
d^4x\frac{\sqrt{-g}}{16\pi}\Big{[}\Omega(\phi)R-\frac{1}{2}g^{\mu
\nu}\partial_{\mu}\phi\partial_{\nu}\phi-U(\phi)\Big{]}+S_m(\psi_m,g_{\mu
\nu})\, ,
\end{equation}
and by making the following conformal transformation,
\begin{equation}\label{ta1higgsintro}
\tilde{g}_{\mu \nu}=A^{-2}g_{\mu \nu}\,
,\,\,\,A(\phi)=\Omega^{-1/2}(\phi)\, ,
\end{equation}
we obtain the Einstein frame scalar field action,
\begin{equation}\label{ta5higgsintro}
\mathcal{S}=\int
d^4x\sqrt{-\tilde{g}}\Big{(}\frac{\tilde{R}}{16\pi}-\frac{1}{2}
\tilde{g}_{\mu \nu}\partial^{\mu}\varphi
\partial^{\nu}\varphi-\frac{V(\varphi)}{16\pi}\Big{)}+S_m(\psi_m,A^2(\varphi)g_{\mu
\nu})\, ,
\end{equation}
where $\varphi$ denotes the Einstein frame canonical scalar field
in the Einstein frame canonical scalar field, with its potential
in the Einstein frame $V(\varphi)$ being related to the Jordan
frame potential $U(\phi)$ as follows,
\begin{equation}\label{potentialns1intro}
V(\varphi)=\frac{U(\phi)}{\Omega^2}\, .
\end{equation}
We also define the useful function $\alpha(\varphi)$,
\begin{equation}\label{alphaofvarphigeneraldefintro}
\alpha(\varphi)=\frac{d \ln A(\varphi)}{d \varphi}\, ,
\end{equation}
which will enter in the final expressions of the TOV equations,
along with the scalar potential $V(\varphi)$ and the function
$A(\varphi)=\Omega^{-1/2}(\phi)$.
For the description of spacetime around and in static NSs, we
shall use the following spherically symmetric metric,
\begin{equation}\label{tov1intro}
ds^2=-e^{\nu(r)}dt^2+\frac{dr^2}{1-\frac{2
m(r)}{r}}+r^2(d\theta^2+\sin^2\theta d\phi^2)\, ,
\end{equation}
where the function $m(r)$ describes the NS gravitational mass and
$r$ denotes the circumferential radius. One of the aims of solving
the TOV equations is to determine in a numerical way the two
functions $\nu(r)$ and $\frac{1}{1-\frac{2 m(r)}{r}}$. Let us
describe in brief the physics of the NS for the scalar-tensor
theory at hand. The metric function $\nu(r)$ has a non-zero value
at the center of the NS, and beyond the NS surface, contrary to
the GR case, the metric is no longer matched to the Schwarzschild
metric, because the functions $\nu(r)$ and $m(r)$ have additional
contributions from the scalar field. Thus in the present case, the
matching of the spherically symmetric metric (\ref{tov1intro})
with the Schwarzschild metric will only be performed at numerical
infinity. The latter will be appropriately determined during the
numerical manipulation of the TOV equations, by checking which
values of the radial parameter $r$ optimize the scalar field
solutions at large distance beyond the surface of the star. The
TOV equations for the Einstein frame scalar field theory for the
metric spherically symmetric metric (\ref{tov1intro}) take the
following form,
\begin{equation}\label{tov2intro}
\frac{d m}{dr}=4\pi r^2
A^4(\varphi)\varepsilon+\frac{r}{2}(r-2m(r))\omega^2+4\pi
r^2V(\varphi)\, ,
\end{equation}
\begin{equation}\label{tov3intro}
\frac{d\nu}{dr}=r\omega^2+\frac{2}{r(r-2m(r))}\Big{[}4\pi
A^4(\varphi)r^3P-4\pi V(\varphi)
r^3\Big{]}+\frac{2m(r)}{r(r-2m(r))}\, ,
\end{equation}
\begin{equation}\label{tov4intro}
\frac{d\omega}{dr}=\frac{4\pi r
A^4(\varphi)}{r-2m(r)}\Big{(}\alpha(\varphi)(\epsilon-3P)+
r\omega(\epsilon-P)\Big{)}-\frac{2\omega
(r-m(r))}{r(r-2m(r))}+\frac{8\pi \omega r^2 V(\varphi)+r\frac{d
V(\varphi)}{d \varphi}}{r-2 m(r)}\, ,
\end{equation}
\begin{equation}\label{tov5intro}
\frac{dP}{dr}=-(\epsilon+P)\Big{[}\frac{1}{2}\frac{d\nu}{dr}+\alpha
(\varphi)\omega\Big{]}\, ,
\end{equation}
\begin{equation}\label{tov5newfinalintro}
\omega=\frac{d \varphi}{dr}\, ,
\end{equation}
where the definition of the function $\alpha (\varphi)$ is given
in Eq. (\ref{alphaofvarphigeneraldefintro}). Also the pressure and
the energy density $P$ and $\epsilon$ denote Jordan frame
quantities, and we shall keep these until the end of the
calculations. Now for the solution of the TOV equations, we need
to solve these numerically for the interior and the exterior of
the NS, independently, and we shall solve the following initial
conditions,
\begin{equation}\label{tov8intro}
P(0)=P_c\, ,\,\,\,m(0)=0\, , \,\,\,\nu(0)\, ,=-\nu_c\, ,
\,\,\,\varphi(0)=\varphi_c\, ,\,\,\, \omega (0)=0\, .
\end{equation}
Note that the values $\nu_c$ and $\varphi_c$ are initially
arbitrary and their exact correct value will be determined by
using a double shooting method which will determine which values
$\nu_c$ and $\varphi_c$ yield the correct behavior of the scalar
field at numerical infinity.

Regarding the EoS, we shall use nine distinct EoSs in order to
describe the nuclear matter inside the NS. We shall adopt a
piecewise polytropic approach \cite{Read:2008iy,Read:2009yp} for
all the EoS, in which the low-density part can be one of the
following EoSs: the SLy \cite{Douchin:2001sv} which is a potential
method EoS, the AP3-AP4 \cite{Akmal:1998cf} which is a variational
method EoS, the WFF1 \cite{Wiringa:1988tp} which is also a
variational method EoS, the ENG \cite{Engvik:1995gn} and the MPA1
\cite{Muther:1987xaa} which are relativistic EoSs, the MS1 and
MS1b \cite{Mueller:1996pm} which are relativistic mean field
theory EoSs, with the MS1b being identical to the MS1 with a low
symmetry energy of 25$\,$MeV and finally the APR EoS
\cite{Akmal:1997ft}. As we shall demonstrate, the MPA1 EoS leads
to remarkable results regarding the compatibility of all the
attractor models with all the constraints. The MPA1 is a
relativistic EoS based on relativistic Brueckner-Hartree-Fock
calculations. We need to note that all the EoS we mentioned above
and which we shall consider, lead to a subluminal maximum speed of
sound speed, save for the WFF1. But as we will show, the WFF1 is
excluded for most of the models, when the observational data are
confronted.

The use of a piecewise polytropic EoS is highly motivated by our
lack of knowledge of the actual relation between the pressure and
the baryon mass density beyond the nuclear density and moreover,
it is still uncertain of what matter is composed of in the core of
NS. What we know though is that the Fermi energy of the matter
particles that compose the NSs is much higher than the temperature
of NSs. The piecewise polytropic EoS is constructed by using
phenomenological data on nuclear matter, this is why it is
considered to be a complete EoS. In principle, the NSs temperature
is much lower compared to the Fermi energy of the particles that
constitute NSs, thus in general NS matter can be described by a
single-parameter polytropic EoS that may accurately describe cold
matter when densities higher than the nuclear density are
considered. However, the uncertainty of the EoS problem always
emerges, which is of course higher as the NS central density is
increased. The pressure considered as a function of the NS
baryonic mass density is in general ill-defined and it is unknown
to at least one order of magnitude, when densities above the
nuclear density are considered. In addition to this issue, the
actual nature of the matter phase at the core of NS is unknown, is
it made of quarks, hyperons? Nobody can answer these questions
rigorously. Thus by using a parameterized EoS engineered to
function at high densities can serve as an optimal scientific
choice for the description of the EoS of NSs. The piecewise
polytropic EoS is exactly this type of EoS. For the construction
of the piecewise polytropic EoS, theoretical and observational
astrophysical constraints are considered, including the very
important causality constraint \cite{Read:2008iy,Read:2009yp}. The
piecewise polytropic EoS is constructed by using a low density
part with density $\rho$ satisfying $\rho<\rho_0$ with $\rho_0$
being the nuclear saturation density, and by construction a well
known tabulated EoS is used, like the SLy or other EoSs. Also a
large density part composes the piecewise polytropic EoS. In fact,
the actual differences between the NS physical quantities, like
the Jordan frame mass and radius, using the piecewise polytropic
EoSs and their ordinary counterparts is of the order
$\mathcal{O}(0.1)\%$ in the context of GR, but still the piecewise
polytropic equation of state approach is more complete from a
phenomenological standpoint for the reasons we discussed above. We
need to note though that there exist studies in the literature
that for medium or low-mass neutron stars, the error of piecewise
polytropic EoSs with ordinary polytropic EoSs can be of the order
$\sim \mathcal{O}(10\%)$ for the Jordan frame mass $M$ and of the
order $\sim \mathcal{O}(2\%)$ for the Jordan frame radius $R$
\cite{OBoyle:2020qvf}. Also alternative spectral EoS studies exist
\cite{Lindblom:2010bb}, which do not yield high errors at lower
masses. Later on we shall present a table in which we shall
compare the Jordan frame masses and radii of NSs with polytropic
and piecewise polytropic EoSs choosing the MPA1 EoS.

In order to maintain the article self-contained, we shall briefly
describe the piecewise polytropic EoS, which is constructed by
using a low-density part describing the crust, with $\rho<\rho_0$,
and $\rho_0$ being obtained by matching the high density part with
the low-density part. The piecewise polytropic EoS also contains
two high densities $\rho_1 = 10^{14.7}{\rm g/cm^3}$ and $\rho_2=
10^{15.0}{\rm g/cm^3}$, and the intermediate pressures and
densities $\rho_{i-1} \leq \rho \leq \rho_i$ satisfy the following
relation,
\begin{equation}\label{pp1}
P = K_i\rho^{\Gamma_i}\, .
\end{equation}
Also we need to require the continuity constraint in each
intermediate patch of the piecewise polytropic EoS,
\begin{equation}\label{pp2}
P(\rho_i) = K_i\rho^{\Gamma_i} = K_{i+1}\rho^{\Gamma_{i+1}}\, .
\end{equation}
From the continuity constraint, by employing Eq. (\ref{pp2}) the
parameters $K_2$ and $K_3$ are derived, given the parameters $K_1,
\Gamma_1, \Gamma_2, \Gamma_3$, which are basically determined by
the low-density part EoS. Each distinct EoS has a different set of
parameters $K_1$ (or equivalently an initial pressure $p_1$),
$\Gamma_1$, $\Gamma_2$, and $\Gamma_3$. The energy density of the
piecewise polytropic EoS as a function of the pressure can be
obtained by using,
\begin{equation}\label{pp3}
d\frac{\epsilon}{\rho} = - P d\frac{1}{\rho}\, ,
\end{equation}
considering always a barotropic fluid, and due to the continuity
equation we get,
\begin{equation}\label{pp4}
\epsilon(\rho) = (1+\alpha_i)\rho +
\frac{K_i}{\Gamma_i-1}\rho^{\Gamma_i}\, ,
\end{equation}
which is valid for $\Gamma_i \neq 1$, and furthermore the
parameter $\alpha_i$ is,
\begin{equation}\label{pp5}
\alpha_i = \frac{\epsilon(\rho_{i-1})}{\rho_{i-1}} -1 -
\frac{K_i}{\Gamma_i-1}\rho_{i-1}^{\Gamma_i-1}\, .
\end{equation}
Having discussed the EoSs we shall use, now let us focus on the
gravitational mass of the static NSs, which is essentially what we
shall calculate numerically when we shall solve the TOV equations.
We shall consider the ADM mass, and the solution of the TOV
equations shall yield the Einstein frame ADM mass, but in the end
we shall transform it to the Jordan frame counterpart. The Jordan
frame ADM is basically what is relevant for weak field
approximation Keplerian orbits. Let us present the Jordan frame
ADM mass expressed in terms of the Einstein frame mass, so let us
introduce $K_E$ and $K_J$ which in Geometrized units are defined
as follows,
\begin{equation}\label{hE}
\mathcal{K}_E=1-\frac{2 m}{r_E}\, ,
\end{equation}
\begin{equation}\label{hE}
\mathcal{K}_J=1-\frac{2  m_J}{r_J}\, .
\end{equation}
The quantities $\mathcal{K}_E$ and $\mathcal{K}_J$ are related as
follows,
\begin{equation}\label{hehjrelation}
\mathcal{K}_J=A^{-2}\mathcal{K}_E\, ,
\end{equation}
and the radii of the NSs in the Jordan and Einstein frames are
related as follows,
\begin{equation}\label{radiiconftrans}
r_J=A r_E\, ,
\end{equation}
and the ADM mass of the NS in the Jordan frame is defined as
follows,
\begin{equation}\label{jordaframemass1}
M_J=\lim_{r\to \infty}\frac{r_J}{2}\left(1-\mathcal{K}_J \right)
\, ,
\end{equation}
and the Einstein frame counterpart is,
\begin{equation}\label{einsteiframemass1}
M_E=\lim_{r\to \infty}\frac{r_E}{2}\left(1-\mathcal{K}_E \right)
\, .
\end{equation}
By taking the asymptotic limit of Eq. (\ref{hehjrelation}), we get
the following formula,
\begin{equation}\label{asymptotich}
\mathcal{K}_J(r_E)=\left(1+\alpha(\varphi(r_E))\frac{d \varphi}{d
r}r_E \right)^2\mathcal{K}_E(\varphi(r_E))\, ,
\end{equation}
where $r_E$ denotes the Einstein frame radius at distances far
away from the surface of the NS, and furthermore $\frac{d\varphi
}{dr}=\frac{d\varphi }{dr}\Big{|}_{r=r_E}$. Using Eqs.
(\ref{hE})-(\ref{asymptotich}) we obtain the final relation of the
Jordan frame ADM mass,
\begin{equation}\label{jordanframeADMmassfinal}
M_J=A(\varphi(r_E))\left(M_E-\frac{r_E^{2}}{2}\alpha
(\varphi(r_E))\frac{d\varphi
}{dr}\left(2+\alpha(\varphi(r_E))r_E\frac{d \varphi}{dr}
\right)\left(1-\frac{2 M_E}{r_E} \right) \right)\, ,
\end{equation}
where recall that $\frac{d\varphi }{dr}=\frac{d\varphi
}{dr}\Big{|}_{r=r_E}$. Hereafter we shall denote the Jordan frame
ADM mass $M_J$ as $M=M_J$ and the central part of our calculation
will be the determination of this mass given in Eq.
(\ref{jordanframeADMmassfinal}).
\begin{figure}[h!]
\centering
\includegraphics[width=18pc]{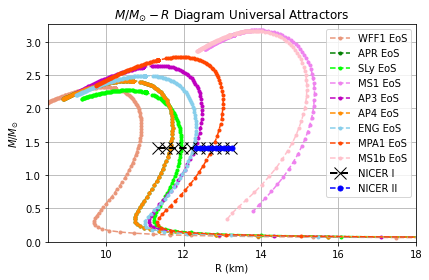}
\includegraphics[width=18pc]{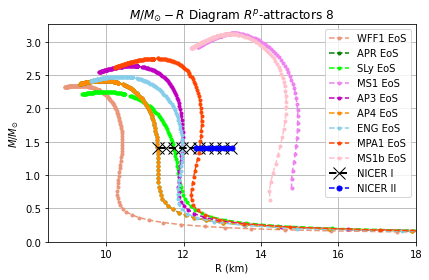}
\includegraphics[width=18pc]{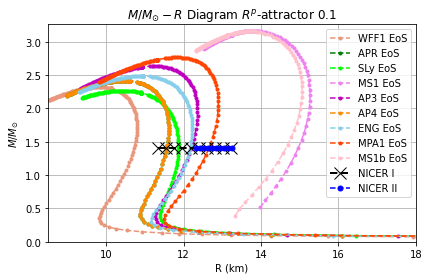}
\includegraphics[width=18pc]{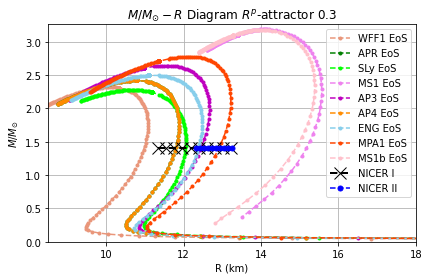}
\includegraphics[width=18pc]{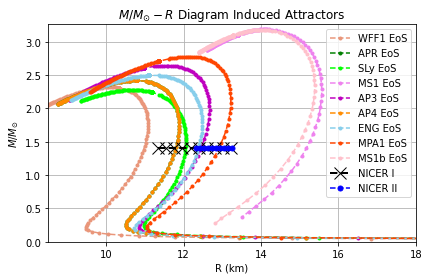}
\includegraphics[width=18pc]{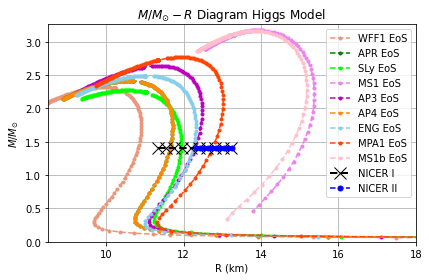}
\caption{The $M-R$ graphs for the universal attractors and the
$R^p$ attractors (three distinct model values), the induced
inflation and Higgs inflation for the EoSs WFF1, SLy, APR, MS1,
AP3, AP4, ENG, MPA1, MS1b versus the NICER I \cite{Miller:2021qha}
and NICER II \cite{Ecker:2022dlg} constraints.} \label{plot1}
\end{figure}
Regarding the radius of the neutron star, we need to evaluate the
Jordan frame one $R$ from the Einstein frame one $R_s$, which are
related by,
\begin{equation}\label{radiussurface}
R=A(\varphi(R_s))\, R_s\, ,
\end{equation}
and the Einstein frame NS mass is determined by $P(R_s)=0$. In the
following we shall numerically calculate the Jordan frame masses
and the radii of the NS for several inflationary attractor
potentials and we shall construct the $M-R$ graphs. In the
following subsections we shall present in brief the inflationary
models we shall consider in this article.

\subsection{Inflation and NSs Physics with $a$-attractors}

The $a$-attractors
\cite{alpha0,alpha1,alpha2,alpha3,alpha4,alpha5,alpha6,alpha7,alpha7a,alpha8,alpha9,alpha10,alpha11,alpha12,alpha13,alpha14,alpha15,alpha16,alpha17,alpha18,alpha19,alpha20,alpha21,alpha22,alpha23,alpha24,alpha25,alpha26,alpha27,alpha28,alpha29,alpha30,alpha31,alpha32,alpha33,alpha34,alpha35,alpha36,alpha37}
are perhaps the most popular class of inflationary models, because
most of the well-known inflationary models compatible with the
latest Planck data \cite{Akrami:2018odb}, fall under this
category. These models produce identical observational indices of
inflation with the Starobinsky model and with the Higgs inflation,
although they originate from distinct Jordan frame theories. The
analysis of NSs physics for some EoSs which will be studied in
this work, can be found in Ref. \cite{Odintsov:2021qbq}. Using the
notation of the previous sections, the $a$-attractors correspond
to the following choice of the kinetic term in the Jordan frame,
\begin{equation}\label{omegaphialphaattractors}
\omega (\phi)=\frac{1}{4\xi}\frac{1}{f}\Big{(}\frac{d
f}{d\phi}\Big{)}^2\, ,
\end{equation}
and furthermore the scalar field potential in the Jordan frame is,
\begin{equation}\label{alphaattractpotensalphaattractors}
U(\phi)=\mathcal{U}_0f^2\left(1-\frac{1}{f} \right)^{2n}\, .
\end{equation}
Then we get,
\begin{equation}\label{finalvarphiphialphaattractors}
\frac{d \varphi }{d
\phi}=\sqrt{\frac{3a}{16\pi}}\frac{1}{f}\Big{(}\frac{d
f}{d\phi}\Big{)}\, ,
\end{equation}
where $a$ is equal to,
\begin{equation}\label{alphaprmalphaattractors}
a=1+\frac{1}{6\xi}\, .
\end{equation}
Upon integration of Eq. (\ref{finalvarphiphialphaattractors}) we
obtain,
\begin{equation}\label{fvarphialphaattractors}
f=e^{\sqrt{\frac{16\pi}{3a}}\varphi}\, ,
\end{equation}
which remarkably holds true irrespectively of the actual form of
the function $f$. The Jordan and Einstein frame potentials are
related by,
\begin{equation}\label{antegeiaalphaattractors}
V(\varphi)=\frac{U(\phi)}{f^2}\, ,
\end{equation}
thus from Eqs. (\ref{alphaattractpotensalphaattractors}) and
(\ref{fvarphialphaattractors}), the Einstein frame scalar field
potential is obtained,
\begin{equation}\label{einsteinframepotentialfinalalphaattractors}
V(\varphi)=\mathcal{U}_0\left(1-e^{-\sqrt{\frac{16\pi}{3a}}\varphi}
\right)^{2n}\, ,
\end{equation}
and we shall use the above for NSs physics for the $a$-attractor
models, in Geometrized units. The conformal factor $A(\phi)$ in
terms of $\varphi$ reads,
\begin{equation}\label{Aofpvarphiprofinalalphaattractors}
A(\varphi)=e^{\alpha \varphi}\, ,
\end{equation}
with $\alpha$ in the case at hand being,
\begin{equation}\label{alphaofphialphaattractors}
\alpha=-\frac{1}{2}\sqrt{\frac{16\pi}{3 a}}\, ,
\end{equation}
and therefore the function  $\alpha(\varphi)$ defined in Eq.
(\ref{alphaofvarphigeneraldefintro}) reads, hence in the case at
hand,
\begin{equation}\label{alphaofphifinalintermsofvarphialphaattractors}
a(\varphi)=\alpha=-\frac{1}{2}\sqrt{\frac{16\pi}{3 a}}\, ,
\end{equation}
and accordingly, the Einstein frame scalar field potential in
terms of the parameter $\alpha$ reads,
\begin{equation}\label{einsteinframepotentialfinalalphaattractors}
V(\varphi)=\mathcal{U}_0\left(1-e^{2\alpha \varphi} \right)^{2n}\,
,
\end{equation}
which will be used for solving the TOV equations later on. One
important feature of inflationary theories, when studying NSs,
usually not taken into account in the existing literature, is the
fact that the values of $\mathcal{U}_0$ are not freely chosen, but
these are constrained by the inflationary theories. The correct
value of $\mathcal{U}_0$ which must be taken into account, can be
found by directly matching the inflationary theory with the
corresponding one in Geometrized units, used in NSs physics. Then
we get that $\mathcal{U}_0$ appearing in Eq.
(\ref{einsteinframepotentialfinalalphaattractors}) is
$\mathcal{U}_0=16\pi \mathcal{V}_0$. Now $\mathcal{V}_0$ is
severely constrained by the inflationary theory, so let us develop
the inflationary theory and thereafter we shall obtain the actual
value of $\mathcal{U}_0$ to be used in the numerical study of the
TOV equations.
\begin{figure}[h!]
\centering
\includegraphics[width=18pc]{higgsalleos.png}
\includegraphics[width=18pc]{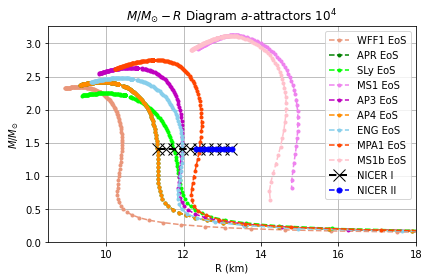}
\includegraphics[width=18pc]{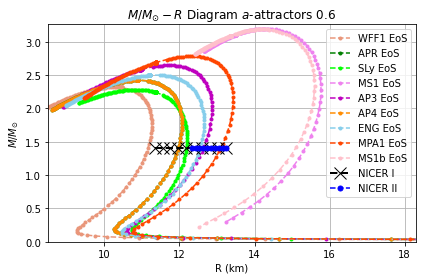}
\includegraphics[width=18pc]{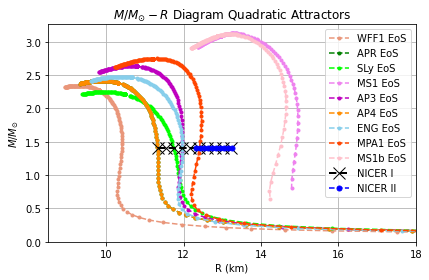}
\caption{The $M-R$ graphs for the universal attractors and the
$R^p$ attractors (three distinct model values), the induced
inflation and Higgs inflation for the EoSs WFF1, SLy, APR, MS1,
AP3, AP4, ENG, MPA1, MS1b versus the NICER I \cite{Miller:2021qha}
and NICER II \cite{Ecker:2022dlg} constraints.} \label{plot2}
\end{figure}
The $a$-attractors have the following Jordan frame inflationary
potential in natural units,
\begin{equation}\label{alphaattractorsjordanpotalphaattractors}
U(\phi)=V_0f^2\Big{(}1-\frac{1}{f}\Big{)}^{2n}\, ,
\end{equation}
and the corresponding Einstein frame scalar potential is,
\begin{equation}\label{einsteinframepotalphaattractors}
V(\varphi)=\mathcal{V}_0\Big{(}1-\frac{1}{f}\Big{)}^{2n}=\tilde{V}_0M_p^4\Big{(}1-\frac{1}{f}\Big{)}^{2n}\,
,
\end{equation}
with $\mathcal{V}_0=\tilde{V}_0M_p^4$ in natural units, and also
$\tilde{V}_0=\frac{V_0}{4}$, therefore $\tilde{V}_0$ is
essentially dimensionless in natural units and furthermore
$\mathcal{V}_0$ has mass dimensions $[m]^4$. The scalar potentials
of the form (\ref{alphaattractorsjordanpotalphaattractors}) are
widely known as $E$-models potentials, and for $n=2$ one obtains
the $a$-attractor potentials. For all the values that the
parameter $a$ can take, the $a$-attractors and $E$-models
remarkably yield the same observational indices of inflation,
namely the same scalar spectral index $n_s$ and the same
tensor-to-scalar ratio $r$ \cite{alpha3}. For our analysis we
shall be interested in values $a\sim \mathcal{O}(10)$ and also we
shall consider values of the order $a\sim \mathcal{O}(10^4)$, and
the inflationary theory is still viable. For small values of $a$,
the observational indices acquire the following form at leading
order in terms of the $e$-foldings number $N$,
\begin{equation}\label{spectralindexsmallalpha}
n_s\simeq 1-\frac{2}{N}\, ,\,\,\,r=\frac{12a}{N^2}\, ,
\end{equation}
while for large values of $a$ ($a\gg \frac{8N}{3}$), the
observational indices take the form \cite{alpha3},
\begin{equation}\label{spectralindexinfalpha}
n_s=1-\frac{2}{N},\,\,\, r=\frac{8}{N}\, .
\end{equation}
Now let us demonstrate the allowed values of $\tilde{V}_0$ given
in Eq. (\ref{einsteinframepotalphaattractors}), which eventually
will determine the allowed values of $\mathcal{U}_0$ given in
(\ref{einsteinframepotentialfinalalphaattractors}). The actual
values of $\tilde{V}_0$ are constrained by the amplitude of the
scalar perturbations for a minimally coupled scalar field, denoted
as $\Delta_s^2$, which is defined to be,
\begin{equation}\label{scalaramp}
\Delta_s^2=\frac{1}{24\pi^2}\frac{V(\varphi_f)}{M_p^4}\frac{1}{\epsilon(\varphi_f)}\,
,
\end{equation}
the values of which are constrained to be \cite{Akrami:2018odb},
\begin{equation}\label{scalarampconst}
\Delta_s^2=2.2\times 10^{-9}\, ,
\end{equation}
where $\varphi_f$ denotes the scalar field value at the end of the
inflationary era, and furthermore $\epsilon$ denotes the first
slow-roll index of inflation. For the case at hand,
\begin{equation}\label{tilde}
\mathcal{V}_0=\tilde{V}_0M_p^4\sim 9.6\times 10^{-11}\, M_p^4\, .
\end{equation}
Thus, in view of the fact that $M_p=1/\sqrt{8 \pi}$ in Geometrized
units, we have for $\mathcal{U}_0$,
\begin{equation}\label{U0constraintalphaattractors}
\mathcal{U}_0=7.62094\times 10^{-12}\, .
\end{equation}
This is the actual value of $\mathcal{U}_0$ which we shall use in
Geometrized units.
\begin{figure}[h!]
\centering
\includegraphics[width=18pc]{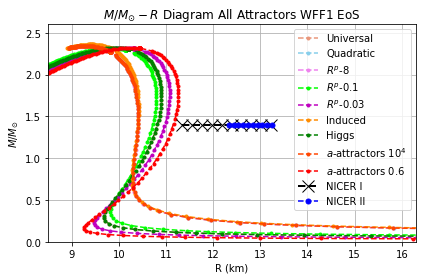}
\includegraphics[width=18pc]{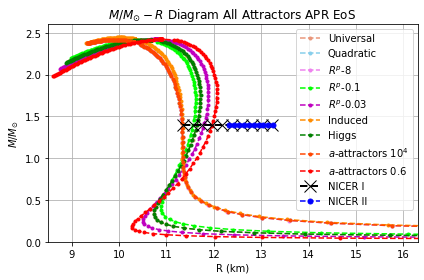}
\includegraphics[width=18pc]{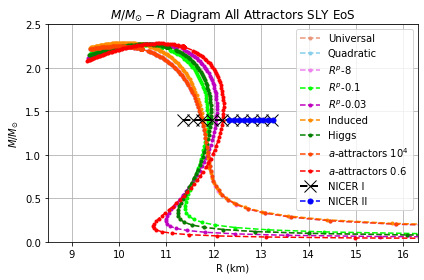}
\includegraphics[width=18pc]{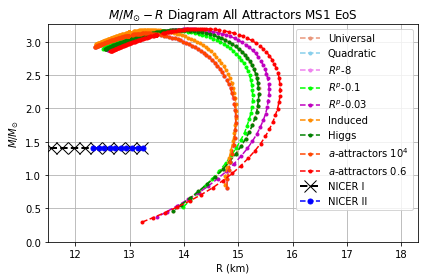}
\caption{The $M-R$ graphs for the universal attractors, the $R^p$
attractors (three distinct model values), the induced inflation,
the quadratic inflation, the Higgs inflation and the
$a$-attractors (two distinct model values) for the EoSs WFF1, SLy,
APR, MS1 versus the NICER I \cite{Miller:2021qha} and NICER II
\cite{Ecker:2022dlg} constraints.} \label{plot3}
\end{figure}

\subsection{Inflation and NSs Physics with  Quadratic and Induced Inflation Attractors}

Now let us consider another mainstream class of inflationary
attractor potentials, namely the quadratic and induced inflation
attractors class which again belong to a wider class of attractors
\cite{alpha0,alpha1,alpha2,alpha3,alpha4,alpha5,alpha6,alpha7,alpha7a,alpha8,alpha9,alpha10,alpha11,alpha12,alpha13,alpha14,alpha15,alpha16,alpha17,alpha18,alpha19,alpha20,alpha21,alpha22,alpha23,alpha24,alpha25,alpha26,alpha27,alpha28,alpha29,alpha30,alpha31,alpha32,alpha33,alpha34,alpha35,alpha36,alpha37}.
These models in the Einstein frame lead to the same minimally
coupled scalar field theory, with similar inflationary properties,
although originating from distinct non-minimally coupled Jordan
frame theories. The analysis of NSs in this kind of theories, for
some of the EoSs used in this article, can be found in Ref.
\cite{submitted}. For both models, the Jordan frame non-minimally
coupled theory is,
\begin{equation}\label{c1quadinduced}
\mathcal{S}_J=\int
d^4x\sqrt{-g}\Big{[}\Omega(\phi)R-\frac{1}{2}g^{\mu
\nu}\partial_{\mu}\phi\partial_{\nu}\phi-U(\phi)\Big{]}+S_m(g_{\mu
\nu},\psi_m)\, ,
\end{equation}
and note that in the case of induced inflation and quadratic
attractors, the kinetic term of the scalar field in the Jordan
frame is canonical. The Jordan frame functions $\Omega(\phi)$ and
$U(\phi)$ have the form \cite{alpha0},
\begin{equation}\label{nnminimalandpotentialquadinduced}
\Omega(\phi)=\left(1+\xi f(\phi)
\right),\,\,\,U(\phi)=M_p^4\lambda \left(\Omega(\phi)-1
\right)^2\, .
\end{equation}
The free parameters $\lambda$, $\xi$ and the non-minimal coupling
function $f (\phi)$, have no mass dimensions in natural units. The
values of the parameter $\xi$, to which we shall refer to as
non-minimal coupling, at strong and weak coupling limit yield the
two distinct models, the induced inflation and the quadratic
attractors. The strong coupling theory $\xi\gg 1$ corresponds to
the induced inflation theory, while the weak coupling theory
$\xi\ll 1$ yields the quadratic attractors. For the induced
inflation attractors, the non-minimal coupling function $\Omega
(\phi)$ and the scalar field potential are,
\begin{equation}\label{inducedpotjordanquadinduced}
\Omega (\phi)=\xi f(\phi),\,\,\,U(\phi)=\lambda M_p^4\xi^2
f(\phi)^2\, .
\end{equation}
The Einstein frame scalar field potential $V(\varphi)$ is,
\begin{equation}\label{c13quadinduced}
V(\varphi)=V_s\left(1-e^{-\sqrt{\frac{2}{3}}\frac{\varphi}{M_p}}
\right)^2\, ,
\end{equation}
with $V_s=\frac{M_p^4\lambda}{\xi^2}$, which is constrained to be,
\begin{equation}\label{tilde}
V_s\sim 9.6\times 10^{-11}\, M_p^4\, ,
\end{equation}
and in effect,  the free parameters $\lambda$ and $\xi$ of the
induced inflation, are chosen so that $\frac{\lambda}{\xi^2}\sim
10^{-11}$, hence for $\xi\sim 10^5$ and for $\lambda \sim 1$ a
viable inflationary era is produced. Furthermore, the non-minimal
coupling function $\Omega (\varphi)$ is,
\begin{equation}\label{nonmimimalinducedquadinduced}
\Omega (\varphi)=e^{\sqrt{\frac{2}{3}}\frac{\varphi}{M_p}}\, .
\end{equation}
The observational indices for inflation at leading order in terms
of the $e$-foldings number read,
\begin{equation}\label{spectralindexsmallalphaquadinduced}
n_s=1-\frac{2}{N}\, ,\,\,\,r=\frac{12}{N^2}\, .
\end{equation}
As it can be seen, the inflationary phenomenology of the induced
inflation attractors is indistinguishable from the Starobinsky and
Higgs inflation model. It is notable that for the induced
inflation attractors, one does not even need to define the
functional form of  the Jordan frame function $f(\phi)$.

Regarding the quadratic attractors, the weak coupling limit of the
same Jordan frame theory, the function $\Omega (\varphi)$ and the
Einstein frame scalar potential are at leading order
\cite{alpha0},
\begin{equation}\label{quadraticfunctionsquadinduced}
\Omega(\varphi )=\xi\left (\xi^{-1}+\frac{g_1}{M_p}\varphi
\right),\,\,\,V(\varphi)=\lambda \frac{g_1^2}{M_p^2}\varphi^2\, ,
\end{equation}
where $g_1\ll \xi^{-1}$, and $g_1$ denotes the expansion
parameter. For the quadratic attractors, the observational indices
of inflation at leading order in the $e$-foldings number $N$ take
the form,
\begin{equation}\label{spectralindexlargelalphaquadinduced}
n_s=1-\frac{2}{N}\, ,\,\,\,r=\frac{8}{N}\, ,
\end{equation}
and note that this result does not depend on the functional form
of $f(\phi)$. From the Einstein frame potential, due to the
constraints on the amplitude of the scalar perturbations, the
parameter $\lambda$ must satisfy $\lambda g_1^2\sim 10^{-11}$ and
we can easily satisfy this by using $\lambda \sim \mathcal{O}(1)$
and $g_1\sim \mathcal{O}(10^{-5})$, while the parameter $\xi$ can
be chosen to be $\xi \sim {10^{-2}}$.

Regarding the NSs analysis, for the induced inflation we have
\begin{equation}\label{inducedAquadinduced}
 A(\varphi)=e^{-\frac{1}{2}\sqrt{\frac{2}{3}}\varphi}\, ,
\end{equation}
and $\alpha (\phi)$ is,
\begin{equation}\label{alphaofphifinalintermsofvarphiquadinduced}
a(\varphi)=-\frac{1}{2}\sqrt{\frac{2}{3}}\, ,
\end{equation}
while the scalar potential in the Einstein frame is,
\begin{equation}\label{indastreinsteinquadinduced}
V(\varphi)=V_s\left(1-e^{-\sqrt{\frac{2}{3}}\varphi} \right)^2\, .
\end{equation}
Regarding the quadratic attractors, the Einstein frame scalar
potential is,
\begin{equation}\label{indastreinsteinquadinduced}
V(\varphi)=\lambda g_1^2\varphi^2\, ,
\end{equation}
and  $A(\varphi)$ is,
\begin{equation}\label{inducedAquadinduced}
 A(\varphi)=\left(1+\xi g_1\varphi \right)^{-1/2}\, ,
\end{equation}
while $\alpha (\varphi)$ is
\begin{equation}\label{alphaofphifinalintermsofvarphiquadinduced}
a(\varphi)=-\frac{1}{2}\frac{\xi g_1}{1+\xi g_1 \varphi}\, .
\end{equation}
Let us recapitulate the values of the free parameters for the two
inflationary attractors, and for the induced inflation case, we
must choose $\xi\sim \mathcal{O}(10^{5})$, $\lambda\sim
\mathcal{O}(1)$ while for the quadratic attractors we choose
$\xi\sim \mathcal{O}(10^{-2})$, $\lambda\sim \mathcal{O}(1)$ and
$g_1\sim \mathcal{O}(10^{-4})$. Usually, in the astrophysics
literature, these parameters would be chosen freely, but these are
not free parameters and are severely constrained by the
corresponding inflationary theories.
\begin{figure}[h!]
\centering
\includegraphics[width=18pc]{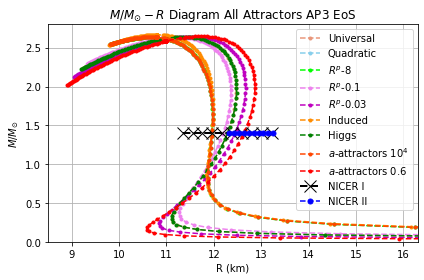}
\includegraphics[width=18pc]{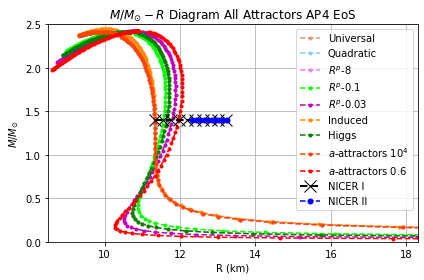}
\includegraphics[width=18pc]{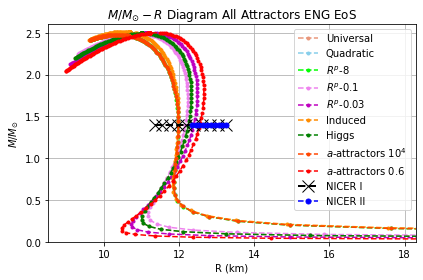}
\includegraphics[width=18pc]{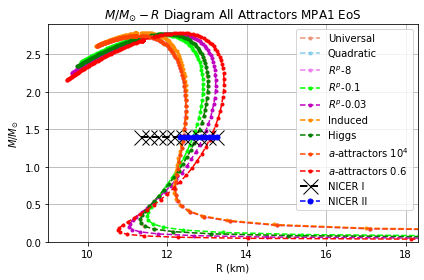}
\includegraphics[width=18pc]{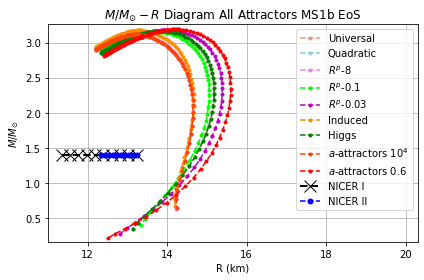}
\caption{The $M-R$ graphs for the universal attractors, the $R^p$
attractors (three distinct model values), the induced inflation,
the quadratic inflation, the Higgs inflation and the
$a$-attractors (two distinct model values) for the EoSs AP3, AP4,
ENG, MPA1, MS1b versus the NICER I \cite{Miller:2021qha} and NICER
II \cite{Ecker:2022dlg} constraints.} \label{plot4}
\end{figure}

\subsection{Inflation and NSs Physics with $R^p$-attractors}

The $R^p$ attractors is a special class of inflationary attractors
which were developed in Ref. \cite{Odintsov:2022bpg}, and are
based on the following Einstein frame scalar potential,
\begin{equation}\label{rpinfinirpattractors}
V(\varphi)= V_0\,M_p^4e^{-2\sqrt{\frac{2}{3}}\kappa
\varphi}\left(e^{\sqrt{\frac{2}{3}}\kappa \varphi}
-1\right)^{\frac{p}{p-1}}\, ,
\end{equation}
In the Jordan frame, the $R^p$ attractors correspond to the
following $F(R)$ gravity,
\begin{equation}\label{JordanframeFRrpattractors}
F(R)=R+\beta R^p\, ,
\end{equation}
with $\beta$ being a free parameter with mass dimensions $[\beta
]=[m]^{2-2p}$ in natural units. The analysis of NSs in this kind
of theories, for some of the EoSs used in this article, can be
found in Ref. \cite{Oikonomou:2023dgu}. The terminology attractors
for the $R^p$ attractors is justified when considering the Jordan
frame non-minimally coupled theory that these originate from. The
action in the Jordan frame of such non-minimally coupled scalar
theory is,
\begin{equation}\label{phijordanframeactionrpattractors}
\mathcal{S}_J=\int
d^4x\left(\frac{\Omega(\phi)}{2\kappa^2}R-\frac{\omega
(\phi)}{2}g^{\mu \nu} \partial_{\mu}\phi \partial_{\nu}\phi
-V_J(\phi)\right)\, ,
\end{equation}
with the coupling function having the general form
$\Omega(\phi)=1+\xi f(\phi)$ and $\xi$, $f(\phi)$ are arbitrary
dimensionless coupling and dimensionless function respectively.
The Jordan frame scalar theory of the $R^p$ attractors correspond
to the following scalar field potential,
\begin{equation}\label{potentialJordanframerpattractors}
V_J(\phi)=V_0\left(\Omega(\phi)-1 \right)^{\frac{p}{p-1}}\, ,
\end{equation}
and the kinetic term $\omega (\phi)$ is,
\begin{equation}\label{kinetictermfunctionrpattractors}
\omega (\phi)=\frac{1}{4\xi}\frac{\left(\frac{d \Omega (\phi)}{d
\phi}\right)^2}{\Omega(\phi)}\, .
\end{equation}
Thus the terminology attractors is justified, since multiple
theories with arbitrary functions $f(\phi)$ correspond to the same
scalar theory in the Einstein frame, with action,
\begin{equation}\label{alphaact}
\mathcal{S}_E=\sqrt{-\tilde{g}}\left(\frac{\tilde{R}}{2\kappa^2}-\tilde{g}^{\mu
\nu}\partial_{\mu}\varphi
\partial_{\nu}\varphi-V(\varphi)\right)\,
,
\end{equation}
The scalar potentials in the Jordan and Einstein frames are
related in the following way,
\begin{equation}\label{jordaneinsteinframepotentialrpattractors}
V(\varphi)=\Omega^{-2}(\phi)V_J(\phi)\, .
\end{equation}
Note that, the general relation between the Jordan and Einstein
frame scalar fields is,
\begin{equation}\label{generalrelationbetweenscalarrpattractors}
\left( \frac{d \varphi}{d
\phi}\right)^{2}=\frac{3}{2}\frac{\left(\frac{d \Omega (\phi)}{d
\phi}\right)^2}{\Omega(\phi)}+\frac{\omega (\phi)}{\Omega(\phi)}\,
,
\end{equation}
therefore, for the $R^p$ attractors case, with the kinetic term
function $\omega (\phi)$ being given in Eq.
(\ref{kinetictermfunctionrpattractors}), we have,
\begin{equation}\label{omegaphigeneralrpattractors}
\Omega (\phi)=e^{\sqrt{\frac{2}{3\alpha}}\varphi}\, ,
\end{equation}
where $\alpha$ is,
\begin{equation}\label{alphageneraldefinitionrpattractors}
\alpha=1+\frac{1}{6\xi}\, .
\end{equation}
Upon substitution of  Eq. (\ref{omegaphigeneralrpattractors}) in
Eq. (\ref{jordaneinsteinframepotentialrpattractors}) we get the
$R^p$-attractor potential. Moreover, the phenomenologically
important case $\alpha=1$ is obtained when $\xi\to \infty$, or
analogously when $\Omega (\phi)\ll \frac{3}{2} \frac{\left(\frac{d
\Omega (\phi)}{d \phi}\right)^2}{\omega(\phi)} $. The $R^p$
attractors theories result to a viable inflationary phenomenology
\cite{Odintsov:2022bpg}, with the observational indices of
inflation being,
\begin{equation}\label{spectralnsexplicitrpattractors}
n_s=\frac{\left(3 \alpha +(3 \alpha -2) p^2+(8-6 \alpha )
p-8\right) e^{2 \sqrt{\frac{2}{3}} \sqrt{\frac{1}{\alpha }} \kappa
\varphi }-2 (p-1) (-3 \alpha +(3 \alpha -2) p+8)
e^{\sqrt{\frac{2}{3}} \sqrt{\frac{1}{\alpha }} \kappa  \varphi
}+(3 \alpha -8) (p-1)^2}{3 \alpha  (p-1)^2
\left(e^{\sqrt{\frac{2}{3}} \sqrt{\frac{1}{\alpha }} \kappa
\varphi }-1\right)^2}\, ,
\end{equation}
\begin{equation}\label{tensotoscalarasfunctionofscalarfieldrpattractors}
r=\frac{16 \left((p-2) e^{\sqrt{\frac{2}{3}} \sqrt{\frac{1}{\alpha
}} \kappa  \varphi }-2 p+2\right)^2}{3 \alpha  (p-1)^2
\left(e^{\sqrt{\frac{2}{3}} \sqrt{\frac{1}{\alpha }} \kappa
\varphi }-1\right)^2}\, .
\end{equation}
and the parameter $V_0$ in the potential is constrained as
follows,
\begin{equation}\label{tilderpattractors}
V_0\sim 9.6\times 10^{-11}\, .
\end{equation}
In this article, we shall consider two limiting cases of $\alpha$,
namely $\alpha\sim 0.1$ and $\alpha=10^8$, with the latter,
covering the strong coupling limit of the theory. Also the
parameter $p$ will be assumed to take values in $1.91\leq p \leq
1.99$, which are the most relevant for NS studies, see Ref.
\cite{Oikonomou:2023dgu} for details. Let us present the relevant
functions for the TOV equations study, the function $A(\varphi)$
for the $R^p$ attractors is,
\begin{equation}\label{inducedArpattractors}
 A(\varphi)=e^{-\frac{1}{2}\sqrt{\frac{2}{3\alpha }}\varphi}\, ,
\end{equation}
hence $\alpha (\phi)$ reads,
\begin{equation}\label{alphaofphifinalintermsofvarphirpattractors}
a(\varphi)=-\frac{1}{2}\sqrt{\frac{2}{3 \alpha }}\, .
\end{equation}
In addition, the Einstein frame scalar potential is,
\begin{equation}\label{rpinfinialphaupdaterpattractors}
V(\varphi)= V_0\,e^{-2\sqrt{\frac{2}{3 \alpha}}
\varphi}\left(e^{\sqrt{\frac{2}{3\alpha }} \varphi}
-1\right)^{\frac{p}{p-1}}\, ,
\end{equation}
and expressed in Geometrized units, the $V_0$ constraint is
\begin{equation}\label{constraintonv0rpattractors}
V_0\simeq 7.62\times 10^{-12}\, .
\end{equation}
\begin{table}[h!]
  \begin{center}
    \caption{\emph{\textbf{Maximum Masses for Inflationary Attractors NSs in the Mass Gap Region.}}}
    \label{tablemaxmasses}
    \begin{tabular}{|r|r|r|r|r|r|}
     \hline
      \textbf{Model}   & \textbf{MPA1 EoS} & \textbf{MS1b EoS} &
      \textbf{AP3 EoS} & \textbf{ENG EoS} & \textbf{MS1 EoS}
      \\  \hline
      \textbf{Universal Attractors $M_{MAX}$} & $M_{MPA1}= 2.771\,M_{\odot}$ & $M_{MS1b}= 3.167\, M_{\odot}$ & $M_{AP3}= 2.638\,
M_{\odot}$ & $M_{ENG}= x $ & $M_{MS1}= 3.174\,M_{\odot}$
\\  \hline
\textbf{$R^p$-8 Attractors $M_{MAX}$} &
$M_{MPA1}=2.749\,M_{\odot}$ & $M_{MS1b}= 3.118\, M_{\odot}$ &
$M_{AP3}= 2.6359\, M_{\odot}$ & $M_{ENG}= x$ &
$M_{MS1}=3.126\,M_{\odot}$
\\  \hline
\textbf{$R^p$-$10^{-1}$ Attractors $M_{MAX}$} & $M_{MPA1}=
2.765\,M_{\odot}$ & $M_{MS1b}= 3.158\, M_{\odot}$ & $M_{AP3}=
2.6357\, M_{\odot}$ & $M_{ENG}= x$ & $M_{MS1}= 3.166\,M_{\odot}$
\\  \hline
\textbf{$R^p$-$10^{-3}$ Attractors $M_{MAX}$} & $M_{MPA1}=
2.778\,M_{\odot}$ & $M_{MS1b}= 3.179\, M_{\odot}$ & $M_{AP3}=
2.643\, M_{\odot}$ & $M_{ENG}= x$ & $M_{MS1}= 3.186\,M_{\odot}$
\\  \hline
\textbf{Induced Attractors $M_{MAX}$} & $M_{MPA1}=
2.788\,M_{\odot}$ & $M_{MS1b}= 3.174\, M_{\odot}$ & $M_{AP3}=
2.659\, M_{\odot}$ & $M_{ENG}= 2.51\,M_{\odot}$ & $M_{MS1}=
3.182\,M_{\odot}$
\\  \hline
\textbf{Higgs Model $M_{MAX}$} & $M_{MPA1}= 2.771\,M_{\odot}$ &
$M_{MS1b}= 3.167\, M_{\odot}$ & $M_{AP3}= 2.638\, M_{\odot}$ &
$M_{ENG}= x$ & $M_{MS1}=3.175\,M_{\odot}$
\\  \hline
\textbf{$\alpha$-$10^4$ Attractors $M_{MAX}$} & $M_{MPA1}=
2.749\,M_{\odot}$ & $M_{MS1b}= 3.117\, M_{\odot}$ & $M_{AP3}=
2.636\, M_{\odot}$ & $M_{ENG}= x$ & $M_{MS1}= 3.126\,M_{\odot}$
\\  \hline
\textbf{$\alpha$-$0.6$  Attractors $M_{MAX}$} & $M_{MPA1}=
2.784\,M_{\odot}$ & $M_{MS1b}= 3.189\, M_{\odot}$ & $M_{AP3}=
2.647\, M_{\odot}$ & $M_{ENG}= 2.500\,M_{\odot}$ & $M_{MS1}=
3.197\,M_{\odot}$
\\  \hline
\textbf{Quadratic Attractors $M_{MAX}$} & $M_{MPA1}=
2.749\,M_{\odot}$ & $M_{MS1b}= 3.117\, M_{\odot}$ & $M_{AP3}=
2.636\, M_{\odot}$ & $M_{ENG}= x$ & $M_{MS1}= 3.126\,M_{\odot}$
\\  \hline
    \end{tabular}
  \end{center}
\end{table}

\subsection{Inflation and NSs Physics with Higgs Model}

The Higgs inflation scalar model is very popular in cosmological
contexts \cite{Mishra:2018dtg}, and the NSs study in this kind of
inflationary potentials was developed in Ref.
\cite{Odintsov:2021nqa}. The Jordan frame action for the Higgs
model is,
\begin{equation}\label{tahiggsmodel}
\mathcal{S}=\int
d^4x\frac{\sqrt{-g}}{16\pi}\Big{[}f(\phi)R-\frac{1}{2}g^{\mu
\nu}\partial_{\mu}\phi\partial_{\nu}\phi-U(\phi)\Big{]}+S_m(\psi_m,g_{\mu
\nu})\, ,
\end{equation}
where $f(\phi)$ and $U(\phi)$ for the Higgs model are,
\begin{equation}\label{fofphihiggsmodel}
f(\phi)=1+\xi \phi^2\, ,
\end{equation}
\begin{equation}\label{jordanframepothiggshiggsmodel}
U(\phi)=\lambda \phi^4\, .
\end{equation}
For the Higgs model, the function $A(\phi)$ is,
\begin{equation}\label{ta2higgshiggsmodel}
A(\phi)=f^{-1/2}(\phi)\, ,
\end{equation}
and in view of Eq. (\ref{fofphihiggsmodel}) we get,
\begin{equation}\label{ta2higgsinihiggsmodel}
A(\phi)=\left(1+\xi \phi^2 \right)^{-1/2}\, .
\end{equation}
The scalar potential reads,
\begin{equation}\label{ta3higgshiggsmodel}
V(\phi)=\frac{U(\phi)}{f^2(\phi)}\, ,
\end{equation}
and expressed in terms of $\phi$ we have,
\begin{equation}\label{potinitialhiggsmodel}
V(\phi)=\frac{\lambda\phi^4}{\left(1+\xi \phi^2 \right)^2}\, .
\end{equation}
The Higgs model in cosmological contexts is obtained for,
\begin{equation}\label{approxmain1higgsmodel}
\xi^2\phi^2\gg 1\, ,
\end{equation}
and at the same time when,
\begin{equation}\label{approxmain2higgsmodel}
\xi^2\phi^2\gg \xi \phi^2\, ,
\end{equation}
with Eq. (\ref{approxmain2higgsmodel}) being valid for $\xi\gg 1$.
Then we get,
\begin{equation}\label{final1higgsmodel}
\frac{d \varphi }{d \phi}\simeq
\frac{\sqrt{12}}{\sqrt{16\pi}}\frac{\xi\phi}{1+\xi\phi^2}=\frac{\sqrt{12}}{2\sqrt{16\pi}}\frac{f'(\phi)}{f(\phi)}\,
,
\end{equation}
and by integrating, Eq. (\ref{final1higgsmodel}) the $\varphi$ and
$\phi$ relation is obtained,
\begin{equation}\label{finalvarphiphirelationhiggsmodel}
\varphi=\frac{\sqrt{12}}{2\sqrt{16\pi}}\ln \left( f(\phi)
\right)=\frac{\sqrt{12}}{2\sqrt{16\pi}}\ln \left( 1+\xi
\phi^2\right)\, ,
\end{equation}
therefore,
\begin{equation}\label{oneplusxiphisquarehiggsmodel}
1+\xi\phi^2=e^{\frac{2\sqrt{16\pi}}{\sqrt{12}}\, \varphi}\, .
\end{equation}
Thus by using Eq. (\ref{ta2higgsinihiggsmodel}), the function
$A(\phi)$ can be expressed in terms of $\varphi$,
\begin{equation}\label{Aofpvarphiprofinalhiggsmodel}
A(\varphi)=e^{\alpha \varphi}\, ,
\end{equation}
with $\alpha$ being,
\begin{equation}\label{alphaofphihiggsmodel}
\alpha=-2\sqrt{\frac{\pi}{3}}\, ,
\end{equation}
and furthermore,
\begin{equation}\label{alphaofphifinalintermsofvarphihiggsmodel}
a(\varphi)=\alpha=-2\sqrt{\frac{\pi}{3}}\, .
\end{equation}
Finally, the Einstein frame Higgs potential is,
\begin{equation}\label{derpotentialfinalofvarphihiggsmodel}
V'(\varphi)\simeq \frac{4 \alpha \lambda}{\xi^2}e^{2\alpha
\varphi}\left(e^{2\alpha \varphi}-1 \right)\, .
\end{equation}
The constraints on the scalar amplitude of curvature
perturbations, are in the Higgs model the following,
\begin{equation}\label{inflconstrainthiggsmodel}
\frac{\lambda M_p^4}{4\xi^2}\sim 9.6\times 10^{-11}M_p^4\, ,
\end{equation}
and in Geometrized units we have,
\begin{equation}\label{constraintfinal}
\frac{\lambda}{\xi^2}\sim 16 \pi \times \left(1.51982\times
10^{-13}\right)\, ,
\end{equation}
therefore for $\lambda=0.1$ we have  $\xi \sim 11.455\times 10^4$.

\subsection{Inflation and NSs Physics with Universal Attractors}

The NSs study of universal attractor potentials was developed in
Ref. \cite{Oikonomou:2021iid}, and in this case, the function
$f(\phi)$ has the form,
\begin{equation}\label{fjordanuniuniversalattractors}
f(\phi)=\frac{M_p^2}{2}\left(1-\xi \phi^2 \right)\, ,
\end{equation}
with $\xi$ being a positive coupling. The Jordan frame scalar
potential is,
\begin{equation}\label{unipotentialjordanuniversalattractors}
U(\phi)=U_0f^2(\phi)\left(\frac{\phi}{M_p}\right)^{2n}\, ,
\end{equation}
where $n>0$. For the universal attractors we have,
\begin{equation}\label{c6universalattractors}
\Omega^2=\frac{2}{M_p^2}f(\phi)\, ,
\end{equation}
and the Einstein frame action is,
\begin{equation}\label{c12universalattractors}
\mathcal{S}_E=\int
d^4x\sqrt{-\tilde{g}}\Big{[}\frac{M_p^2}{2}\tilde{R}-\frac{\zeta
(\phi)}{2} \tilde{g}^{\mu \nu }\tilde{\partial}_{\mu}\phi
\tilde{\partial}_{\nu}\phi-V(\phi)\Big{]}+S_m(\Omega^{-2}\tilde{g}_{\mu
\nu},\psi_m)\, ,
\end{equation}
with,
\begin{equation}\label{c14universalattractors}
\zeta
(\phi)=\frac{M_p^2}{2}\Big{(}\frac{3\Big{(}\frac{df}{d\phi}\Big{)}^2}{f^2}+\frac{2\omega(\phi)}{f}\Big{)}\,
,
\end{equation}
while the potentials in the two frames are related as follows,
\begin{equation}\label{c13universalattractors}
V(\phi)=\frac{U(\phi)}{\Omega^4}\, ,
\end{equation}
with,
\begin{equation}\label{unipotentialjordanuniversalattractors}
V(\phi)=\frac{U_0 M_p^4}{4}\left(\frac{\phi}{M_p}\right)^{2n}\, .
\end{equation}
In the case of universal attractors, the observational indices of
inflation are again,
\begin{equation}\label{spectralindexsmallalpha}
n_s=1-\frac{2}{N}\, ,\,\,\,r=\frac{12}{N^2}\, ,
\end{equation}
just like in the Higgs inflation and several other attractors
cases.
\begin{table}[h!]
  \begin{center}
    \caption{\emph{\textbf{Inflationary Attractors vs CSI for NS Masses $M\sim 2M_{\odot}$, $R_{2M_{\odot}}=12.11^{+1.11}_{-1.23}\,$km, for the SLy, APR, WFF1, MS1 and AP3 EoSs.}}}
    \label{tablecsi2}
    \begin{tabular}{|r|r|r|r|r|r|}
     \hline
      \textbf{Model}   & \textbf{SLy EoS} & \textbf{APR EoS} & \textbf{WFF1
      EoS} & \textbf{MS1 EoS} & \textbf{AP3 EoS}
      \\  \hline
      \textbf{Universal Attractors Radii} & $R_{SLy}= 11.64\,$Km & $R_{APR}=  11.63\,$Km & $R_{WFF1}=
      x$
       & $R_{MS1}= x$ & $R_{AP3}= 12.46\,$Km
\\  \hline
\textbf{$R^p$-8 Attractors Radii} & $R_{SLy}= 11.15\,$Km &
$R_{APR}=  11.08\,$Km & $R_{WFF1}=
      x$
       & $R_{MS1}= x$ & $R_{AP3}= 11.89\,$Km
\\  \hline
\textbf{$R^p$-$10^{-1}$ Attractors Radii} & $R_{SLy}= 11.54\,$Km &
$R_{APR}=  11.52\,$Km & $R_{WFF1}=
      x$
       & $R_{MS1}= x$ & $R_{AP3}= 12.35\,$Km
\\  \hline
\textbf{$R^p$-$10^{-3}$ Attractors Radii} & $R_{SLy}= 11.78\,$Km &
$R_{APR}=  11.785\,$Km & $R_{WFF1}=
      11.01\,$Km
       & $R_{MS1}= x$ & $R_{AP3}= 12.66\,$Km
\\  \hline
\textbf{Induced Attractors Radii} & $R_{SLy}= 11.17\,$Km &
$R_{APR}=  11.08\,$Km & $R_{WFF1}=
      x$
       & $R_{MS1}= x$ & $R_{AP3}= 11.94\,$Km
\\  \hline
\textbf{Higgs Model Radii} & $R_{SLy}= 11.64\,$Km & $R_{APR}=
11.60\,$Km & $R_{WFF1}=
      x$
       & $R_{MS1}= x$ & $R_{AP3}= 12.47\,$Km
\\  \hline
\textbf{$\alpha$-$10^4$ Attractors Radii} & $R_{SLy}= 11.17\,$Km &
$R_{APR}=  11.00\,$Km & $R_{WFF1}=
      x$
       & $R_{MS1}= x$ & $R_{AP3}= 11.86\,$Km
\\  \hline
\textbf{$\alpha$-$10^{-1}$  Attractors Radii} & $R_{SLy}=
11.96\,$Km & $R_{APR}=  11.96\,$Km & $R_{WFF1}=
      11.16\,$Km
       & $R_{MS1}= x$ & $R_{AP3}= 12.85\,$Km
\\  \hline
\textbf{Quadratic Attractors Radii} & $R_{SLy}= 11.17\,$Km &
$R_{APR}=  11.06\,$Km & $R_{WFF1}=
      x$
       & $R_{MS1}= x$ & $R_{AP3}= 11.89\,$Km
\\  \hline
    \end{tabular}
  \end{center}
\end{table}

\begin{table}[h!]
  \begin{center}
    \caption{\emph{\textbf{Inflationary Attractors vs CSI for NS Masses $M\sim
2M_{\odot}$, $R_{2M_{\odot}}=12.11^{+1.11}_{-1.23}\,$km, for the
AP4, ENG, MPA1 and MS1b.}}}
    \label{tablecsi22}
    \begin{tabular}{|r|r|r|r|r|}
     \hline
      \textbf{Model}   & \textbf{AP4 EoS} & \textbf{ENG EoS} &
      \textbf{MPA1 EoS} & \textbf{MS1b EoS}
      \\  \hline
      \textbf{Universal Attractors Radii} & $R_{AP4}= 11.60\,$Km & $R_{ENG}=  12.22\,$Km & $R_{MPA1}=
      13.01\,$Km
       & $R_{MS1b}= x$
\\  \hline
\textbf{$R^p$-8 Attractors Radii} & $R_{AP4}= 11.05\,$Km &
$R_{ENG}=  11.74 \,$Km & $R_{MPA1}=
      112.44\,$Km
       & $R_{MS1b}= x$
\\  \hline
\textbf{$R^p$-$10^{-1}$ Attractors Radii} & $R_{AP4}= 11.52\,$Km &
$R_{ENG}=  12.14\,$Km & $R_{MPA1}=12.89
      \,$Km
       & $R_{MS1b}= x$
\\  \hline
\textbf{$R^p$-$10^{-3}$ Attractors Radii} & $R_{AP4}= 11.82\,$Km &
$R_{ENG}=  12.42\,$Km & $R_{MPA1}= 13.21\,$Km
       & $R_{MS1b}= x$
\\  \hline
\textbf{Induced Attractors Radii} & $R_{AP4}= 11.15\,$Km &
$R_{ENG}= 11.82\,$Km & $R_{MPA1}=
      12.46\,$Km
       & $R_{MS1b}= x$
\\  \hline
\textbf{Higgs Model Radii} & $R_{AP4}= 11.63\,$Km & $R_{ENG}=
12.24\,$Km & $R_{MPA1}=
      13.02\,$Km
       & $R_{MS1b}= x$
\\  \hline
\textbf{$\alpha$-$10^4$ Attractors Radii} & $R_{AP4}= 11.08\,$Km &
$R_{ENG}=  11.74\,$Km & $R_{MPA1}=
      12.44\,$Km
       & $R_{MS1b}= x$
\\  \hline
\textbf{$\alpha$-$10^{-1}$  Attractors Radii} & $R_{AP4}=
12.01\,$Km & $R_{ENG}=  12.61\,$Km & $R_{MPA1}=13.42\,$Km
       & $R_{MS1b}= x$
\\  \hline
\textbf{Quadratic Attractors Radii} & $R_{AP4}= 11.08\,$Km &
$R_{ENG}=  11.68\,$Km & $R_{MPA1}=12.44
      \,$Km
       & $R_{MS1b}= x$
\\  \hline
    \end{tabular}
  \end{center}
\end{table}
For the universal attractor models, it is assumed that,
\begin{equation}\label{omegacons1universalattractors}
\Omega(\phi)\ll \frac{3 M_p^2}{2}\Omega'(\phi)\, ,
\end{equation}
and this can be written as follows,
\begin{equation}\label{conditionomega2universalattractors}
1-\frac{\xi \phi^2}{M_p^2}\ll \frac{6 \xi^2 \phi^2}{M_p^2}\, .
\end{equation}
Using,
\begin{equation}\label{c14proequniversalattractors}
\frac{d\varphi}{d\phi}=\frac{\sqrt{1+\frac{\xi
\phi^2}{M_p^2}+\frac{6\xi^2\phi^2}{M_p^2}}}{1+\frac{\xi
\phi^2}{M_p^2}}\, ,
\end{equation}
and also Eq. (\ref{omegacons1universalattractors}), we get,
\begin{equation}\label{finalphioffuniversalattractors}
\varphi=-\sqrt{\frac{3}{2}}M_p\ln \left(1-\frac{\xi \phi^2}{M_p^2}
\right)\, ,
\end{equation}
which can be rewritten,
\begin{equation}\label{finalfofvarphiuniversalattractors}
\frac{\phi^2}{M_p^2}=\frac{1-e^{-\sqrt{\frac{2}{3}}\frac{\varphi}{M_p}}}{\xi}\,
.
\end{equation}
Thus the Einstein frame scalar potential has the final form in
terms of $\varphi$,
\begin{equation}\label{finalpotentialeinsteininflationuniversalattractors}
V(\varphi)=\frac{U_0
M_p^4}{4\xi^n}\left(1-e^{-\sqrt{\frac{2}{3}}\frac{\varphi}{M_p}}\right)^{n}\,
,
\end{equation}
and by setting $V_0=\frac{U_0 M_p^4}{4\xi^n}$, due to the
inflationary constraints on the amplitude of scalar perturbations,
$V_0$ is constrained,
\begin{equation}\label{tildeuniversalattractors}
V_0\sim 9.6\times 10^{-11}\, M_p^4\, .
\end{equation}
which constraints in effect the parameter $\xi$ to be
$\xi=36.2239\times 10^{4}$. Now regarding the functions that enter
in the TOV equations for the universal attractors case, we have,
for $A(\varphi)$,
\begin{equation}\label{conformalfactor}
A(\varphi)=e^{2\sqrt{\frac{\pi}{3}}\varphi}\, ,
\end{equation}
while $a(\varphi)$ reads,
\begin{equation}\label{alphaofphifinalintermsofvarphi}
a(\varphi)=\alpha=2\sqrt{\frac{\pi}{3}}\, .
\end{equation}
\begin{table}[h!]
  \begin{center}
    \caption{\emph{\textbf{Inflationary Attractors vs CSI for NS Masses $M\sim 1.4M_{\odot}$, $R_{1.4M_{\odot}}=12.42^{+0.52}_{-0.99}$, for the SLy, APR, WFF1, MS1 and AP3 EoSs. }}}
    \label{tablecsi14}
    \begin{tabular}{|r|r|r|r|r|r|}
     \hline
      \textbf{Model}   & \textbf{SLy EoS} & \textbf{APR EoS} & \textbf{WFF1
      EoS} & \textbf{MS1 EoS} & \textbf{AP3 EoS}
      \\  \hline
      \textbf{Universal Attractors Radii} & $R_{SLy}= 11.93\,$Km & $R_{APR}=  11.64\,$Km & $R_{WFF1}=
      x$
       & $R_{MS1}= x$ & $R_{AP3}= x$
\\  \hline
\textbf{$R^p$-8 Attractors Radii} & $R_{SLy}= 11.73\,$Km &
$R_{APR}=  x$ & $R_{WFF1}=
      x$
       & $R_{MS1}= x$ & $R_{AP3}= x$
\\  \hline
\textbf{$R^p$-$10^{-1}$ Attractors Radii} & $R_{SLy}= 11.85\,$Km &
$R_{APR}=  11.55\,$Km & $R_{WFF1}=
      x$
       & $R_{MS1}= x$ & $R_{AP3}= 12.18\,$Km
\\  \hline
\textbf{$R^p$-$10^{-3}$ Attractors Radii} & $R_{SLy}= 12.04\,$Km &
$R_{APR}=  11.80\,$Km & $R_{WFF1}=
      x$
       & $R_{MS1}= x$ & $R_{AP3}= 12.43\,$Km
\\  \hline
\textbf{Induced Attractors Radii} & $R_{SLy}= 11.77\,$Km &
$R_{APR}=  x$ & $R_{WFF1}=
      x$
       & $R_{MS1}= x$ & $R_{AP3}= 11.96\,$Km
\\  \hline
\textbf{Higgs Model Radii} & $R_{SLy}= 11.93\,$Km & $R_{APR}=
11.64\,$Km & $R_{WFF1}=
      x$
       & $R_{MS1}= x$ & $R_{AP3}= 12.28\,$Km
\\  \hline
\textbf{$\alpha$-$10^4$ Attractors Radii} & $R_{SLy}= 11.73\,$Km &
$R_{APR}=  x$ & $R_{WFF1}=
      x$
       & $R_{MS1}= x$ & $R_{AP3}= 11.96\,$Km
\\  \hline
\textbf{$\alpha$-$10^{-1}$  Attractors Radii} & $R_{SLy}=
12.18\,$Km & $R_{APR}=  11.95\,$Km & $R_{WFF1}=
      x$
       & $R_{MS1}= x$ & $R_{AP3}= 12.65\,$Km
\\  \hline
\textbf{Quadratic Attractors Radii} & $R_{SLy}= 11.75\,$Km &
$R_{APR}=  11.64\,$Km & $R_{WFF1}=
      x$
       & $R_{MS1}= x$ & $R_{AP3}= 11.96\,$Km
\\  \hline
    \end{tabular}
  \end{center}
\end{table}

\begin{table}[h!]
  \begin{center}
    \caption{\emph{\textbf{Inflationary Attractors vs CSI for NS Masses $M\sim
1.4M_{\odot}$, $R_{1.4M_{\odot}}=12.42^{+0.52}_{-0.99}$, for the
AP4, ENG, MPA1 and MS1b.}}}
    \label{tablecsi142}
    \begin{tabular}{|r|r|r|r|r|}
     \hline
      \textbf{Model}   & \textbf{AP4 EoS} & \textbf{ENG EoS} &
      \textbf{MPA1
      EoS} & \textbf{MS1b EoS}
      \\  \hline
      \textbf{Universal Attractors Radii} & $R_{AP4}= 11.64\,$Km & $R_{ENG}=  12.23\,$Km & $R_{MPA1}=
      12.74\,$Km
       & $R_{MS1b}= x$
\\  \hline
\textbf{$R^p$-8 Attractors Radii} & $R_{AP4}= 11.93\,$Km &
$R_{ENG}=  x $ & $R_{MPA1}=
      12.41\,$Km
       & $R_{MS1b}= x$
\\  \hline
\textbf{$R^p$-$10^{-1}$ Attractors Radii} & $R_{AP4}= 11.55\,$Km &
$R_{ENG}=  12.16\,$Km & $R_{MPA1}=12.64
      \,$Km
       & $R_{MS1b}= x$
\\  \hline
\textbf{$R^p$-$10^{-3}$ Attractors Radii} & $R_{AP4}= 11.80\,$Km &
$R_{ENG}=  12.38\,$Km & $R_{MPA1}= x$
       & $R_{MS1b}= x$
\\  \hline
\textbf{Induced Attractors Radii} & $R_{AP4}= x$ & $R_{ENG}=
11.97\,$Km & $R_{MPA1}=
      12.38\,$Km
       & $R_{MS1b}= x$
\\  \hline
\textbf{Higgs Model Radii} & $R_{AP4}= 11.64\,$Km & $R_{ENG}=
12.23\,$Km & $R_{MPA1}=
      12.74\,$Km
       & $R_{MS1b}= x$
\\  \hline
\textbf{$\alpha$-$10^4$ Attractors Radii} & $R_{AP4}= x$ &
$R_{ENG}=  11.97\,$Km & $R_{MPA1}=
      12.41\,$Km
       & $R_{MS1b}= x$
\\  \hline
\textbf{$\alpha$-$10^{-1}$  Attractors Radii} & $R_{AP4}=
11.95\,$Km & $R_{ENG}=  12.55\,$Km & $R_{MPA1}=x$
       & $R_{MS1b}= x$
\\  \hline
\textbf{Quadratic Attractors Radii} & $R_{AP4}= 11.93\,$Km &
$R_{ENG}=  11.97\,$Km & $R_{MPA1}=12.41
      x$
       & $R_{MS1b}= x$
\\  \hline
    \end{tabular}
  \end{center}
\end{table}


\subsection{Results and Confrontation with the Data}

Let us now present the results of our numerical analysis we
performed using the LSODA technique for solving the TOV equations.
We analyzed all the inflationary attractors we mentioned in the
previous sections, namely, the $a$-attractors for two
characteristic values of the free parameters, the $R^p$-attractors
for three characteristic values of the free parameters, the
universal attractors, the quadratic attractors, the Higgs and
induced inflation cases. We used the following EoSs, as we
mentioned earlier, the SLy \cite{Douchin:2001sv} which is a
potential method EoS, the AP3-AP4 \cite{Akmal:1998cf} which is a
variational method EoS, the WFF1 \cite{Wiringa:1988tp} which is
also a variational method EoS, the ENG \cite{Engvik:1995gn} and
the MPA1 \cite{Muther:1987xaa} which are relativistic EoSs, the
MS1 and MS1b \cite{Mueller:1996pm} which are relativistic mean
field theory EoSs, with the MS1b being identical to the MS1 with a
low symmetry energy of 25$\,$MeV and finally the APR EoS
\cite{Akmal:1997ft}. We constructed the $M-R$ graphs for all these
cases, and we confronted the extracted data on masses and radii of
the resulting static NSs with the three following constraints CSI,
CSII and CSIII, also appearing in Fig. \ref{plotcs}: the CSI
constraint appeared firstly in Ref. \cite{Altiparmak:2022bke} and
indicates that the radius of an $1.4M_{\odot}$ mass NS must be
$R_{1.4M_{\odot}}=12.42^{+0.52}_{-0.99}$ while the radius of a
$2M_{\odot}$ mass NS must be
$R_{2M_{\odot}}=12.11^{+1.11}_{-1.23}\,$km. With regard to CSII,
it appeared firstly in Ref. \cite{Raaijmakers:2021uju} and
indicates that the radius of an $1.4M_{\odot}$ mass NS must be
$R_{1.4M_{\odot}}=12.33^{+0.76}_{-0.81}\,\mathrm{km}$. With regard
to the CSIII constraint, it appeared firstly in Ref.
\cite{Bauswein:2017vtn} and indicates two things: firstly that the
radius of an $1.6M_{\odot}$ mass NS has be larger than
$R_{1.6M_{\odot}}>10.68^{+0.15}_{-0.04}\,$km and secondly, the
radius of a NS corresponding to its maximum mass, must satisfy
$R_{M_{max}}>9.6^{+0.14}_{-0.03}\,$km.
\begin{table}[h!]
  \begin{center}
    \caption{\emph{\textbf{Inflationary Attractors Radii vs CSII for NS Masses $M\sim 1.4M_{\odot}$, $R_{1.4M_{\odot}}=12.33^{+0.76}_{-0.81}\,\mathrm{km}$, for the SLy, APR, WFF1, MS1 and AP3 EoSs. }}}
    \label{tablecsii14}
    \begin{tabular}{|r|r|r|r|r|r|}
     \hline
      \textbf{Model}   & \textbf{SLy EoS} & \textbf{APR EoS} & \textbf{WFF1
      EoS} & \textbf{MS1 EoS} & \textbf{AP3 EoS}
      \\  \hline
      \textbf{Universal Attractors Radii} & $R_{SLy}= 11.93\,$Km & $R_{APR}=  11.64\,$Km & $R_{WFF1}=
      x$
       & $R_{MS1}= x$ & $R_{AP3}= 11.36\,$Km
\\  \hline
\textbf{$R^p$-8 Attractors Radii} & $R_{SLy}= 11.73\,$Km &
$R_{APR}=  x$ & $R_{WFF1}=
      x$
       & $R_{MS1}= x$ & $R_{AP3}= x$
\\  \hline
\textbf{$R^p$-$10^{-1}$ Attractors Radii} & $R_{SLy}= 11.85\,$Km &
$R_{APR}=  11.55\,$Km & $R_{WFF1}=
      x$
       & $R_{MS1}= x$ & $R_{AP3}= 12.18\,$Km
\\  \hline
\textbf{$R^p$-$10^{-3}$ Attractors Radii} & $R_{SLy}= 12.04\,$Km &
$R_{APR}=  11.80\,$Km & $R_{WFF1}=
      x$
       & $R_{MS1}= x$ & $R_{AP3}= 12.43\,$Km
\\  \hline
\textbf{Induced Attractors Radii} & $R_{SLy}= 11.77\,$Km &
$R_{APR}=  x$ & $R_{WFF1}=
      x$
       & $R_{MS1}= x$ & $R_{AP3}= 11.96\,$Km
\\  \hline
\textbf{Higgs Model Radii} & $R_{SLy}= 11.93\,$Km & $R_{APR}=
11.64\,$Km & $R_{WFF1}=
      x$
       & $R_{MS1}= x$ & $R_{AP3}= 12.28\,$Km
\\  \hline
\textbf{$\alpha$-$10^4$ Attractors Radii} & $R_{SLy}= 11.73\,$Km &
$R_{APR}=  x$ & $R_{WFF1}=
      x$
       & $R_{MS1}= x$ & $R_{AP3}= 11.96\,$Km
\\  \hline
\textbf{$\alpha$-$10^{-1}$  Attractors Radii} & $R_{SLy}=
12.18\,$Km & $R_{APR}=  11.95\,$Km & $R_{WFF1}=
      x$
       & $R_{MS1}= x$ & $R_{AP3}= 12.65\,$Km
\\  \hline
\textbf{Quadratic Attractors Radii} & $R_{SLy}= 11.75\,$Km &
$R_{APR}=  11.64\,$Km & $R_{WFF1}=
      x$
       & $R_{MS1}= x$ & $R_{AP3}= 11.96\,$Km
\\  \hline
    \end{tabular}
  \end{center}
\end{table}
\begin{table}[h!]
  \begin{center}
    \caption{\emph{\textbf{Inflationary Attractors vs CSII for NS Masses $M\sim
1.4M_{\odot}$,
$R_{1.4M_{\odot}}=12.33^{+0.76}_{-0.81}\,\mathrm{km}$, for the
AP4, ENG, MPA1 and MS1b.}}}
    \label{tablecsii142}
    \begin{tabular}{|r|r|r|r|r|}
     \hline
      \textbf{Model}   & \textbf{AP4 EoS} & \textbf{ENG EoS} &
      \textbf{MPA1 EoS} & \textbf{MS1b EoS}\\  \hline
      \textbf{Universal Attractors Radii} & $R_{AP4}= 11.64\,$Km & $R_{ENG}=  12.23\,$Km & $R_{MPA1}=
      12.74\,$Km  & $R_{MS1b}= x$
\\  \hline
\textbf{$R^p$-8 Attractors Radii} & $R_{AP4}= 11.93\,$Km &
$R_{ENG}=  x $ & $R_{MPA1}=
      12.41\,$Km
       & $R_{MS1b}= x$
\\  \hline
\textbf{$R^p$-$10^{-1}$ Attractors Radii} & $R_{AP4}= 11.55\,$Km &
$R_{ENG}=  12.16\,$Km & $R_{MPA1}=12.64
      \,$Km
       & $R_{MS1b}= x$
\\  \hline
\textbf{$R^p$-$10^{-3}$ Attractors Radii} & $R_{AP4}= 11.80\,$Km &
$R_{ENG}=  12.38\,$Km & $R_{MPA1}= x$
       & $R_{MS1b}= x$
\\  \hline
\textbf{Induced Attractors Radii} & $R_{AP4}= x$ & $R_{ENG}=
11.97\,$Km & $R_{MPA1}=
      12.38\,$Km
       & $R_{MS1b}= x$
\\  \hline
\textbf{Higgs Model Radii} & $R_{AP4}= 11.64\,$Km & $R_{ENG}=
12.23\,$Km & $R_{MPA1}=
      12.74\,$Km
       & $R_{MS1b}= x$
\\  \hline
\textbf{$\alpha$-$10^4$ Attractors Radii} & $R_{AP4}= x$ &
$R_{ENG}=  11.97\,$Km & $R_{MPA1}=
      12.41\,$Km
       & $R_{MS1b}= x$
\\  \hline
\textbf{$\alpha$-$10^{-1}$  Attractors Radii} & $R_{AP4}=
11.95\,$Km & $R_{ENG}=  12.55\,$Km & $R_{MPA1}=x$
       & $R_{MS1b}= x$
\\  \hline
\textbf{Quadratic Attractors Radii} & $R_{AP4}= 11.93\,$Km &
$R_{ENG}=  11.97\,$Km & $R_{MPA1}=12.41
      x$
       & $R_{MS1b}= x$
\\  \hline
    \end{tabular}
  \end{center}
\end{table}
The TOV equations are solved numerically in order to extract the
Jordan frame masses and radii of the NSs and in order to construct
the corresponding $M-R$ graphs. We used a python-3-based numerical
code which is based on the pyTOV-STT code \cite{niksterg}. Using
the LSODA integration technique, and a double shooting method in
order to find the optimal values of the initial conditions $\nu_c$
and $\varphi_c$ defined in Eq. (\ref{tov8intro}), which render the
scalar field zero at numerical infinity. Let us further elaborate
on the numerical integration technique. As we quoted above, the
method we shall use is the LSODA, which is more powerful compared
to the Runge-Kutta technique, due to the fact that the LSODA
integration method is more appropriate for stiff differential
equation problems. We solved separately the TOV equations for the
interior and the exterior of the star, using approximately 160
values of the central density, and nearly 20000 grid points. The
method is quite reliable because we used a double shooting
technique to determine the accurate values for the $\nu_c$ and
$\varphi_c$ defined in Eq. (\ref{tov8intro}), and specifically we
optimized their values according to the rule that for the optimal
values, the scalar field vanishes at numerical infinity. The
numerical infinity is actually determined by exactly this rule.

For all the $M-R$ graphs we shall plot, we included the NICER I
constraint for $M=1.4M_{\odot}$ NSs which is $90\%$ credible
\cite{Miller:2021qha} and constrains the radius of an
$M=1.4M_{\odot}$ to be $R_{1.4M_{\odot}}=11.34-13.23\,$km. Also
recently, a more refined extension of the NICER constraint was
introduced in \cite{Ecker:2022dlg} by taking into account the
heavy black-widow binary pulsar PSR J0952-0607 with mass
$M=2.35\pm 0.17$ \cite{Romani:2022jhd}. The black widow pulsar PSR
J0952-0607 is the heavier known NSs and we need to note that the
NICER II constraint take this into account. The results of our
analysis are quite interesting, aligned with the NICER II
constraints and indicate that these are satisfied only for EoS
which yield the maximum masses of NSs inside the mass-gap region,
but still lower that the causal 3 solar masses limit. The NICER II
constraints the radius of a $M=1.4M_{\odot}$ to be
$R_{1.4M_{\odot}}=12.33-13.25\,$km. We shall refer to this new
constraint as NICER II for simplicity hereafter. Also in the
following Table \ref{nicerconstraints} we present the two NICER I
and NICER II constraints for reading convenience.
\begin{table}[h!]
  \begin{center}
    \caption{\emph{\textbf{NICER I AND NICER II Constraints for the radius of a $M=1.4M_{\odot}$ NS. }}}
    \label{nicerconstraints}
    \begin{tabular}{|r|r|}
 \hline
  NICER I & $11.34\,\mathrm{km}<R_{1.4M_{\odot}}<13.23\,\mathrm{km}$ \cite{Miller:2021qha} \\
  \hline
  NICER II & $12.33\,\mathrm{km}<R_{1.4M_{\odot}}<13.25\,\mathrm{km}$ \cite{Ecker:2022dlg} \\
  \hline
    \end{tabular}
  \end{center}
\end{table}
The $M-R$ graphs of our analysis are presented in Figs.
\ref{plot1}, \ref{plot2}, \ref{plot3} and \ref{plot4}. In Fig.
\ref{plot1} we present the $M-R$ graphs for the universal
attractors and the $R^p$ attractors (three distinct model values),
the induced inflation and Higgs inflation for the EoSs WFF1, SLy,
APR, MS1, AP3, AP4, ENG, MPA1, MS1b versus the NICER I
\cite{Miller:2021qha} and NICER II \cite{Ecker:2022dlg}
constraints. In Fig. \ref{plot2} we present the $M-R$ graphs for
the universal attractors and the $R^p$ attractors (three distinct
model values), the induced inflation and Higgs inflation for the
EoSs WFF1, SLy, APR, MS1, AP3, AP4, ENG, MPA1, MS1b versus the
NICER I and NICER II constraints. In Fig. \ref{plot3} we present
the $M-R$ graphs for the universal attractors, the $R^p$
attractors (three distinct model values), the induced inflation,
the quadratic inflation, the Higgs inflation and the
$a$-attractors (two distinct model values) for the EoSs WFF1, SLy,
APR, MS1 versus the NICER I and NICER II constraints and finally
in Fig. \ref{plot4}, the $M-R$ graphs are presented for all the
aforementioned attractor models, for the EoSs AP3, AP4, ENG, MPA1,
MS1b, confronted again with the NICER I, II constraints.
\begin{table}[h!]
  \begin{center}
    \caption{\emph{\textbf{Inflationary Attractors vs CSIII for NS Masses $M\sim 1.6M_{\odot}$, $R_{1.6M_{\odot}}>10.68^{+0.15}_{-0.04}\,$km, for the SLy, APR, WFF1, MS1 and AP3 EoSs. }}}
    \label{tablecsiii16}
    \begin{tabular}{|r|r|r|r|r|r|}
     \hline
      \textbf{Model}   & \textbf{SLy EoS} & \textbf{APR EoS} & \textbf{WFF1EoS} & \textbf{MS1 EoS} & \textbf{AP3 EoS}
      \\  \hline
      \textbf{Universal Attractors Radii} & $R_{SLy}= 11.92\,$Km & $R_{APR}=  11.70\,$Km & $R_{WFF1}=
     11.68\,$Km
       & $R_{MS1}= 13.99\,$Km & $R_{AP3}= 12.41\,$Km
\\  \hline
\textbf{$R^p$-8 Attractors Radii} & $R_{SLy}= 11.89\,$Km &
$R_{APR}= 11.28\,$Km & $R_{WFF1}=
      10.40\,$Km
       & $R_{MS1}= 14.93\,$Km & $R_{AP3}= 11.97\,$Km
\\  \hline
\textbf{$R^p$-$10^{-1}$ Attractors Radii} & $R_{SLy}= 11.85\,$Km &
$R_{APR}= 11.61\,$Km & $R_{WFF1}=
      x$
       & $R_{MS1}= 15.11\,$Km & $R_{AP3}= 12.29\,$Km
\\  \hline
\textbf{$R^p$-$10^{-3}$ Attractors Radii} & $R_{SLy}= 11.80\,$Km &
$R_{APR}= 11.59 \,$Km & $R_{WFF1}=
      x$
       & $R_{MS1}= 15.20\,$Km & $R_{AP3}= 12.34\,$Km
\\  \hline
\textbf{Induced Attractors Radii} & $R_{SLy}= 11.68\,$Km &
$R_{APR}= 11.32\,$Km & $R_{WFF1}=
      x$
       & $R_{MS1}= 14.92\,$Km & $R_{AP3}= 11.97\,$Km
\\  \hline
\textbf{Higgs Model Radii} & $R_{SLy}= 11.92\,$Km & $R_{APR}=
11.70\,$Km & $R_{WFF1}=
      10.89\,$Km
       & $R_{MS1}= 14.92\,$Km & $R_{AP3}= 11.93\,$Km
\\  \hline
\textbf{$\alpha$-$10^4$ Attractors Radii} & $R_{SLy}= 11.64\,$Km &
$R_{APR}=  11.29\,$Km & $R_{WFF1}=
      x$
       & $R_{MS1}= 14.94\,$Km & $R_{AP3}= 11.97\,$Km
\\  \hline
\textbf{$\alpha$-$10^{-1}$  Attractors Radii} & $R_{SLy}=
12.20\,$Km & $R_{APR}=  12.04\,$Km & $R_{WFF1}=
      11.23\,$Km
       & $R_{MS1}= 15.49\,$Km & $R_{AP3}= 12.76\,$Km
\\  \hline
\textbf{Quadratic Attractors Radii} & $R_{SLy}= 11.64\,$Km &
$R_{APR}=  11.29\,$Km & $R_{WFF1}=
      x$
       & $R_{MS1}= 14.94\,$Km & $R_{AP3}= 11.97\,$Km
\\  \hline
    \end{tabular}
  \end{center}
\end{table}
\begin{table}[h!]
  \begin{center}
    \caption{\emph{\textbf{Inflationary Attractors vs CSIII for NS Masses $M\sim
1.6M_{\odot}$, $R_{1.6M_{\odot}}>10.68^{+0.15}_{-0.04}\,$km, for
the AP4, ENG, MPA1 and MS1b.}}}
    \label{tablecsiii162}
    \begin{tabular}{|r|r|r|r|r|}
     \hline
      \textbf{Model}   & \textbf{AP4 EoS} & \textbf{ENG EoS} &
      \textbf{MPA1
      EoS} & \textbf{MS1b EoS}
      \\  \hline
      \textbf{Universal Attractors Radii} & $R_{AP4}= 11.70\,$Km & $R_{ENG}=  12.30\,$Km & $R_{MPA1}=
      12.87\,$Km
       & $R_{MS1b}= 14.90\,$Km
\\  \hline
\textbf{$R^p$-8 Attractors Radii} & $R_{AP4}= 11.28\,$Km &
$R_{ENG}=  11.94 \,$Km & $R_{MPA1}=
      12.45\,$Km
       & $R_{MS1b}= 14.59\,$Km
\\  \hline
\textbf{$R^p$-$10^{-1}$ Attractors Radii} & $R_{AP4}= 11.61\,$Km &
$R_{ENG}=  12.21\,$Km & $R_{MPA1}=12.79
      \,$Km
       & $R_{MS1b}= 14.78\,$Km
\\  \hline
\textbf{$R^p$-$10^{-3}$ Attractors Radii} & $R_{AP4}= 11.86\,$Km &
$R_{ENG}=  12.45\,$Km & $R_{MPA1}= 12.15\,$Km
       & $R_{MS1b}= 15.04\,$Km
\\  \hline
\textbf{Induced Attractors Radii} & $R_{AP4}= 11.32\,$Km &
$R_{ENG}= 11.96\,$Km & $R_{MPA1}=
      12.44\,$Km
       & $R_{MS1b}= 14.49\,$Km
\\  \hline
\textbf{Higgs Model Radii} & $R_{AP4}= 11.70\,$Km & $R_{ENG}=
12.30\,$Km & $R_{MPA1}=
      12.87\,$Km
       & $R_{MS1b}= 14.90\,$Km
\\  \hline
\textbf{$\alpha$-$10^4$ Attractors Radii} & $R_{AP4}= 11.28\,$Km &
$R_{ENG}=  11.95\,$Km & $R_{MPA1}=
      12.44\,$Km
       & $R_{MS1b}= 14.55\,$Km
\\  \hline
\textbf{$\alpha$-$10^{-1}$  Attractors Radii} & $R_{AP4}=
10.99\,$Km & $R_{ENG}=  12.63\,$Km & $R_{MPA1}=13.22\,$Km
       & $R_{MS1b}= 15.22\,$Km
\\  \hline
\textbf{Quadratic Attractors Radii} & $R_{AP4}= 11.28\,$Km &
$R_{ENG}=  11.95\,$Km & $R_{MPA1}=12.44
      \,$Km
       & $R_{MS1b}= 14.55\,$Km
\\  \hline
    \end{tabular}
  \end{center}
\end{table}
Also in Table \ref{tablemaxmasses} we present the EoSs and the
inflationary attractor models which yield a maximum NS mass inside
the mass-gap region. In Tables \ref{tablecsi2}-\ref{tablecsi22} we
present the confrontation of the radii of the NSs with the
constraint CSI regarding $M\sim 2M_{\odot}$ NSs. In Tables
\ref{tablecsi14}-\ref{tablecsi142} we present the confrontation of
the radii of the NSs with the constraint CSI regarding $M\sim
1.4M_{\odot}$ NSs. In Tables \ref{tablecsii14}-\ref{tablecsii142}
we present the confrontation of the radii of the NSs with the
constraint CSII regarding $M\sim 1.4M_{\odot}$ NSs. In Tables
\ref{tablecsiii16}-\ref{tablecsiii162} we present the
confrontation of the radii of the NSs with the constraint CSIII
regarding $M\sim 1.6M_{\odot}$ NSs, while in Tables
\ref{tablecsiiimax}-\ref{tablecsiiimax2} we present the
confrontation of the radii of the NSs with the constraint CSIII
regarding maximum mass NSs.

Now let us analyze the extracted data, and the results are deemed
quite interesting. The most important result is that all the
inflationary models are compatible with the NICER I and NICER II,
and all the CSI, CSII, CSIII constraints, for the MPA1 EoS. More
importantly, for this particular EoS, the maximum predicted NS
masses are inside the mass gap region $2.5-5$$M_{\odot}$, but well
below the causal 3 solar masses limit of GR, which is respected.
Recall that the causal EoS limit is very important so let us
briefly recall what the causal EoS limit is, and why our result is
aligned with this extreme GR limit, which also holds true in
modified gravity theories \cite{Astashenok:2021peo}.
\begin{table}[h!]
  \begin{center}
    \caption{\emph{\textbf{Inflationary Attractors Maximum NS Masses and the Corresponding Radii vs CSIII, $R_{M_{max}}>9.6^{+0.14}_{-0.03}\,$km, for the SLy, APR, WFF1, MS1 and AP3 EoSs. }}}
    \label{tablecsiiimax}
    \begin{tabular}{|r|r|r|r|r|r|}
     \hline
      \textbf{Model}   & \textbf{APR EoS} & \textbf{SLy EoS} & \textbf{WFF1
      EoS} & \textbf{MS1 EoS} & \textbf{AP3 EoS}
      \\  \hline
      \textbf{Universal Attractors $M_{max}$} & $M_{APR}= 2.41\,M_{\odot}$ & $M_{SLy}= 2.27\, M_{\odot}$ & $M_{WFF1}= 2.32\,
M_{\odot}$ & $M_{MS1}= 3.17\,M_{\odot}$ & $M_{AP3}=
2.63\,M_{\odot}$
\\  \hline
\textbf{Universal Attractors Radii} & $R_{APR}= 10.54\,$Km &
$R_{SLy}= 10.56\,$Km & $R_{WFF1}=
     9.91\,$Km
       & $R_{MS1}= 13.93\,$Km & $R_{AP3}= 11.35\,$Km
\\  \hline
\textbf{$R^p$-8 Attractors $M_{max}$} & $M_{APR}= 2.41\,M_{\odot}$
& $M_{SLy}= 2.24\, M_{\odot}$ & $M_{WFF1}= 2.34\, M_{\odot}$ &
$M_{MS1}= 3.12\,M_{\odot}$ & $M_{AP3}= 2.63\,M_{\odot}$
\\  \hline
\textbf{$R^p$-8 Attractors Radii} & $R_{APR}= 9.91\,$Km &
$R_{SLy}= 9.96\,$Km& $R_{WFF1}=
      9.28\,$Km
       & $R_{MS1}= 13.30\,$Km & $R_{AP3}= 10.67\,$Km
\\  \hline
\textbf{$R^p$-$10^{-1}$ Attractors $M_{max}$} & $M_{APR}=
2.41\,M_{\odot}$ & $M_{SLy}= 2.26\, M_{\odot}$ & $M_{WFF1}= 2.31\,
M_{\odot}$ & $M_{MS1}= 3.16\,M_{\odot}$ & $M_{AP3}=
2.63\,M_{\odot}$
\\  \hline
\textbf{$R^p$-$10^{-1}$ Attractors Radii} & $R_{APR}= 10.49\,$Km &
$R_{SLy}= 10.43\,$Km & $R_{WFF1}= 9.83\,$Km
       & $R_{MS1}= 13.81\,$Km & $R_{AP3}= 11.24\,$Km
\\  \hline
\textbf{$R^p$-$10^{-3}$ Attractors $M_{max}$} & $M_{APR}=
2.42\,M_{\odot}$ & $M_{SLy}= 2.27\, M_{\odot}$ & $M_{WFF1}= 2.32\,
M_{\odot}$ & $M_{MS1}= 3.18\,M_{\odot}$ & $M_{AP3}=
2.64\,M_{\odot}$
\\  \hline
\textbf{$R^p$-$10^{-3}$ Attractors Radii} & $R_{APR}= 10.77 \,$Km
&  $R_{SLy}= 10.72\,$Km& $R_{WFF1}=
      10.09\,$Km
       & $R_{MS1}= 14.09\,$Km & $R_{AP3}= 11.55\,$Km
\\  \hline
\textbf{Induced Attractors $M_{max}$} & $M_{APR}= 2.44\,M_{\odot}$
& $M_{SLy}= 2.28\, M_{\odot}$ & $M_{WFF1}= 2.36\, M_{\odot}$ &
$M_{MS1}= 3.18\,M_{\odot}$ & $M_{AP3}= 2.66\,M_{\odot}$
\\  \hline
\textbf{Induced Attractors Radii} & $R_{APR}= 10.04\,$Km &
$R_{SLy}= 10.10\,$Km& $R_{WFF1}=
      9.42\,$Km
       & $R_{MS1}= 13.36\,$Km & $R_{AP3}= 10.80\,$Km
\\  \hline
\textbf{Higgs Model $M_{max}$} & $M_{APR}= 2.41\,M_{\odot}$ &
$M_{SLy}= 2.27\, M_{\odot}$ & $M_{WFF1}= 2.32\, M_{\odot}$ &
$M_{MS1}= 3.17\,M_{\odot}$ & $M_{AP3}= 2.63\,M_{\odot}$
\\  \hline
\textbf{Higgs Model Radii} & $R_{APR}= 10.57\,$Km & $R_{SLy}=
10.56\,$Km & $R_{WFF1}=
      9.91\,$Km
       & $R_{MS1}= 13.91\,$Km & $R_{AP3}= 11.35\,$Km
\\  \hline
\textbf{$\alpha$-$10^4$ Attractors $M_{max}$} & $M_{APR}=
2.41\,M_{\odot}$ & $M_{SLy}= 2.24\, M_{\odot}$ & $M_{WFF1}= 2.34\,
M_{\odot}$ & $M_{MS1}= 3.12\,M_{\odot}$ & $M_{AP3}=
2.63\,M_{\odot}$
\\  \hline
\textbf{$\alpha$-$10^4$ Attractors Radii} & $R_{APR}= 9.89\,$Km &
$R_{SLy}= 9.98\,$Km
 & $R_{WFF1}=
      9.29\,$Km
       & $R_{MS1}= 13.31\,$Km & $R_{AP3}= 10.65\,$Km
\\  \hline
\textbf{$\alpha$-$10^{-1}$  Attractors $M_{max}$} & $M_{APR}=
2.42\,M_{\odot}$ & $M_{SLy}= 2.28\, M_{\odot}$ & $M_{WFF1}= 2.32\,
M_{\odot}$ & $M_{MS1}= 3.19\,M_{\odot}$ & $M_{AP3}=
2.64\,M_{\odot}$
\\  \hline
\textbf{$\alpha$-$10^{-1}$  Attractors Radii} & $R_{APR}=
10.89\,$Km &$R_{SLy}= 10.81\,$Km  & $R_{WFF1}=
      10.22\,$Km
       & $R_{MS1}= 14.29\,$Km & $R_{AP3}= 11.72\,$Km
\\  \hline
\textbf{Quadratic Attractors $M_{max}$} & $M_{APR}=
2.41\,M_{\odot}$ & $M_{SLy}= 2.24\, M_{\odot}$ & $M_{WFF1}= 2.34\,
M_{\odot}$ & $M_{MS1}= 3.12\,M_{\odot}$ & $M_{AP3}=
2.63\,M_{\odot}$
\\  \hline
\textbf{Quadratic Attractors Radii} & $R_{APR}=  9.89\,$Km
&$R_{SLy}= 9.96\,$Km
 & $R_{WFF1}=
      9.28\,$Km
       & $R_{MS1}= 13.31\,$Km & $R_{AP3}= 10.67\,$Km
\\  \hline
    \end{tabular}
  \end{center}
\end{table}
\begin{table}[h!]
  \begin{center}
    \caption{\emph{\textbf{Inflationary Attractors Maximum NS Masses and the correspondent vs CSIII, $R_{M_{max}}>9.6^{+0.14}_{-0.03}\,$km, for the AP4, ENG, MPA1 and MS1b }}}
    \label{tablecsiiimax2}
    \begin{tabular}{|r|r|r|r|r|}
     \hline
      \textbf{Model}   & \textbf{AP4 EoS} & \textbf{ENG EoS} &
      \textbf{MPA1
      EoS} & \textbf{MS1b EoS}
      \\  \hline
      \textbf{Universal Attractors $M_{max}$} & $M_{AP4}= 2.41\,M_{\odot}$ & $M_{ENG}= 2.49\, M_{\odot}$ & $M_{MPA1}= 2.77\,
M_{\odot}$ & $M_{MS1b}= 3.16\,M_{\odot}$
\\  \hline
\textbf{Universal Attractors Radii} & $R_{AP4}= 10.54\,$Km &
$R_{ENG}=  11.00\,$Km & $R_{MPA1}=
      11.94\,$Km
       & $R_{MS1b}= 13.83\,$Km
\\  \hline
\textbf{$R^p$-8 Attractors $M_{max}$} & $M_{AP4}= 2.41\,M_{\odot}$
& $M_{ENG}= 2.47\, M_{\odot}$ & $M_{MPA1}= 2.74\, M_{\odot}$ &
$M_{MS1b}= 3.11\,M_{\odot}$
\\  \hline
\textbf{$R^p$-8 Attractors Radii} & $R_{AP4}= 9.91\,$Km &
$R_{ENG}=  10.36 \,$Km & $R_{MPA1}=
      11.33\,$Km
       & $R_{MS1b}= 13.21\,$Km
\\  \hline
\textbf{$R^p$-$10^{-1}$ Attractors $M_{max}$} & $M_{AP4}=
2.41\,M_{\odot}$ & $M_{ENG}= 2.48\, M_{\odot}$ & $M_{MPA1}= 2.76\,
M_{\odot}$ & $M_{MS1b}= 3.15\,M_{\odot}$
\\  \hline
\textbf{$R^p$-$10^{-1}$ Attractors Radii} & $R_{AP4}= 10.49\,$Km &
$R_{ENG}=  10.92\,$Km & $R_{MPA1}=11.82
      \,$Km
       & $R_{MS1b}= 13.73\,$Km
\\  \hline
\textbf{$R^p$-$10^{-3}$ Attractors $M_{max}$} & $M_{AP4}=
2.42\,M_{\odot}$ & $M_{ENG}= 2.49\, M_{\odot}$ & $M_{MPA1}= 2.77\,
M_{\odot}$ & $M_{MS1b}= 3.17\,M_{\odot}$
\\  \hline
\textbf{$R^p$-$10^{-3}$ Attractors Radii} & $R_{AP4}= 10.77\,$Km &
$R_{ENG}=  11.11\,$Km & $R_{MPA1}= 12.15\,$Km
       & $R_{MS1b}= 14.06\,$Km
\\  \hline
\textbf{Induced Attractors $M_{max}$} & $M_{AP4}= 2.44\,M_{\odot}$
& $M_{ENG}= 2.51\, M_{\odot}$ & $M_{MPA1}= 2.78\, M_{\odot}$ &
$M_{MS1b}= 3.17\,M_{\odot}$
\\  \hline
\textbf{Induced Attractors Radii} & $R_{AP4}= 10.04\,$Km &
$R_{ENG}= 10.49\,$Km & $R_{MPA1}=
      11.41\,$Km
       & $R_{MS1b}= 13.29\,$Km
\\  \hline
\textbf{Higgs Model $M_{max}$} & $M_{AP4}= 2.41\,M_{\odot}$ &
$M_{ENG}= 2.49\, M_{\odot}$ & $M_{MPA1}= 2.77\, M_{\odot}$ &
$M_{MS1b}= 3.16\,M_{\odot}$
\\  \hline
\textbf{Higgs Model Radii} & $R_{AP4}= 10.57\,$Km & $R_{ENG}=
11.01\,$Km & $R_{MPA1}=
      11.94\,$Km
       & $R_{MS1b}= 13.83\,$Km
\\  \hline
\textbf{$\alpha$-$10^4$ Attractors $M_{max}$} & $M_{AP4}=
2.41\,M_{\odot}$ & $M_{ENG}= 2.47\, M_{\odot}$ & $M_{MPA1}=
2.342\, M_{\odot}$ & $M_{MS1b}= 2.417\,M_{\odot}$
\\  \hline
\textbf{$\alpha$-$10^4$ Attractors Radii} & $R_{AP4}= 9.89\,$Km &
$R_{ENG}=  10.36\,$Km & $R_{MPA1}=
      12.44\,$Km
       & $R_{MS1b}= 14.55\,$Km
\\  \hline
\textbf{$\alpha$-$10^{-1}$  Attractors $M_{max}$} & $M_{AP4}=
2.42\,M_{\odot}$ & $M_{ENG}= 2.50\, M_{\odot}$ & $M_{MPA1}= 2.78\,
M_{\odot}$ & $M_{MS1B}= 3.18\,M_{\odot}$
\\  \hline
\textbf{$\alpha$-$10^{-1}$  Attractors Radii} & $R_{AP4}=
10.89\,$Km & $R_{ENG}=  11.48\,$Km & $R_{MPA1}=12.33\,$Km
       & $R_{MS1b}= 14.22\,$Km
\\  \hline
\textbf{Quadratic Attractors $M_{max}$} & $M_{AP4}=
2.41\,M_{\odot}$ & $M_{ENG}= 2.47\, M_{\odot}$ & $M_{MPA1}= 2.74\,
M_{\odot}$ & $M_{MS1b}= 3.11\,M_{\odot}$
\\  \hline
\textbf{Quadratic Attractors Radii} & $R_{AP4}= 9.89\,$Km &
$R_{ENG}=  10.36\,$Km & $R_{MPA1}=11.32
      \,$Km
       & $R_{MS1b}= 13.21\,$Km
\\  \hline
    \end{tabular}
  \end{center}
\end{table}
The causality constraint is an important constraint for most
reliable EoSs, which must be respected in order for an EoS to be
considered reliable, In principle, when $dP/d\rho>1$ in natural
units, the EoS can be rendered superluminal, while when $P>\rho$,
the EoS can become ultrabaric. A related question of course is
whether a superluminality predicting EoS in terms of its sound
speed can be incompatible with Lorentz invariance and of course
causality. The answer to this question is no, nevertheless, the
EoSs which become superluminal, do become superluminal for NSs
energy densities for which the NSs are rendered hydrodynamically
unstable \cite{Haensel:2007yy}, and this applies for stiff EoSs
too. If we take into consideration the nuclear matter stability
condition at high densities $\frac{dP}{d\rho}>0$, in conjunction
with the subluminality condition $\frac{dP}{d\rho}\leq 1$ (in
natural units), a very well established GR originating limit for
the maximum mass of static NS is
 the so-called causal maximum mass limit \cite{Rhoades:1974fn,Kalogera:1996ci},
\begin{equation}\label{causalupperbound}
M_{max}^{CL}=3M_{\odot}\sqrt{\frac{5\times
10^{14}g/cm^{3}}{\rho_u}}\, ,
\end{equation}
with $\rho_u$ being the maximum density up to which the EoS is
well-known and the corresponding pressure is $P_u(\rho_u)$. The
causal limit EoS is the following,
\begin{equation}\label{causallimiteos}
P_{sn}(\rho)=P_{u}(\rho_u)+(\rho-\rho_u)c^2\, .
\end{equation}
Another well known bound in astrophysics, also supporting the
causal maximum mass limit, is the bound based on the maximum
baryon mass of a static NS (all NSs with periods $P\succeq 3$ms),
\begin{equation}\label{3solarmasslimit}
M_{max}\leq 3M_{\odot}\, .
\end{equation}
The causal maximum mass of a NSs when rotation is taken into
account is,
\begin{equation}\label{causalrot}
M^{CL,rot}_{max}=3.89M_{\odot}\sqrt{\frac{5\times
10^{14}g/cm^{3}}{\rho_u}}\, .
\end{equation}
In our case it is remarkable that the MPA1 EoS, produces results
compatible with all the constraints for all the inflationary
attractors and it is also compatible with the causal limit of
maximum mass (\ref{causalupperbound}). The only EoSs which produce
maximum masses beyond the causal limit of maximum mass
(\ref{causalupperbound}) are the MS1 and MS1b, and remarkably,
none of the two produces $M-R$ graphs which are compatible with
the constraints. In fact, none of the two is compatible with even
a single constraint. Another important feature of our analysis is
the fact that if the NICER II constraints are taken into account,
which recall that generated based on the heavier PSR J0952-0607,
it seems that compatibility with the data comes hand in hand with
models and EoS which predict maximum masses beyond $2.5$ solar
masses, thus in the mass gap region and at the same time below the
3 solar masses causal limit. This for example becomes true for all
the inflationary models studied, only for the MPA1 EoS, so the
question is, is this EoS so important? Is it possible that this
EoS describes NSs at a fundamental level? The future observations
will show, but for the moment we report this interesting behavior.
Of course there are other EoSs that are compatible with the NICER
II constraint, such as the ENG and AP3, which remarkably also
predict maximum masses inside the mass gap region and below the 3
solar masses causal limit, but in these EoSs, only some
inflationary attractors produce viable results, for example the
Higgs, the $a$-attractors and the $R^p$ attractors, but in the
case of the MPA1 EoS, all the inflationary models are compatible
with the NICER II observations. Another major outcome of this work
is that cosmologically indistinguishable inflationary attractors,
become distinguished in NSs. This feature however seems to be EoS
and model dependent, since for example for the APR EoS, the
quadratic attractors, the $a$-attractors and the $R^p$-8
attractors produce overlapping results. Regarding the constraint
CSI, the WFF1, the MS1 and MS1b EoS are excluded completely (apart
for the some cases of the $a$-attractors and $R^p$-attractors
regarding the WFF1) and the same applies for the CSII constraint.
Regarding the CSIII constraints, only the WFF1 EoS produces
results which are in most cases of inflationary attractors,
incompatible with CSIII. The full results of the incompatibility
of certain inflationary attractor models and EoSs, with the
observational data and the existing constraints, can be found in
the presented tables. We highlighted the most significant outcomes
of this work, and as it seems, the MPA1 EoS seems to play an
important role, while the WFF1, MS1 and MS1b seem to be entirely
out of the equation, regarding a viable description of static NSs.

Before closing, let us discuss an important issue related with the
discrimination of inflationary attractors on NS, the
$\%$-difference between the NS masses for different models of
attractors and the differences in NS masses corresponding to
piecewise polytropic and polytropic EoS. It is vital that the
differences between the masses using piecewise polytropic EoSs and
ordinary polytropic EoS to be lower than the differences between
NS masses corresponding to inflationary attractors. In Table
\ref{reftable} we present the maximum masses of all the
inflationary attractors for the piecewise polytropic MPA1 EoS and
for the ordinary polytropic MPA1 EoS. The difference in the
maximum mass between the different attractors for the piecewise
polytropic MPA1 EoS varies from the minimum value $\sim
\mathcal{O}(0.8\%)$ to the maximum value $\sim
\mathcal{O}(1.27\%)$, while the average differences between the
piecewise polytropic and the ordinary polytropic MPA1 EoS is of
the order $\sim \mathcal{O}(0.1\%)$. Thus this feature may somehow
obscure the discrimination of different attractors in NSs and it
is a reportable feature, because the differences in maximum masses
may be within the limits of errors in determining the mass of the
NS.

\begin{table}[h!]
  \begin{center}
    \caption{\emph{\textbf{Comparison of the Maximum Masses for All the Inflationary Attractors NSs for the Piecewise and Ordinary Polytropic MPA1 EoS .}}}
    \label{reftable}
    \begin{tabular}{|r|r|r|}
     \hline
      \textbf{Model}   & \textbf{MPA1 Piecewise Polytropic EoS} & \textbf{MPA1 Ordinary Polytropic EoS}
      \\  \hline
      \textbf{Universal Attractors $M_{MAX}$} & $M^{pp}_{MPA1}= 2.771\,M_{\odot}$ & $M^{p}_{MPA1}= 2.77433\,M_{\odot}$
\\  \hline
\textbf{$R^p$-8 Attractors $M_{MAX}$} &
$M^{pp}_{MPA1}=2.749\,M_{\odot}$ & $M^{p}_{MPA1}= 2.74571\,
M_{\odot}$
\\  \hline
\textbf{$R^p$-$10^{-1}$ Attractors $M_{MAX}$} & $M^{pp}_{MPA1}=
2.765\,M_{\odot}$ & $M^{p}_{MPA1}= 2.76804\, M_{\odot}$
\\  \hline
\textbf{$R^p$-$10^{-3}$ Attractors $M_{MAX}$} & $M^{pp}_{MPA1}=
2.778\,M_{\odot}$ & $M^{p}_{MPA1}= 2.78078\, M_{\odot}$
\\  \hline
\textbf{Induced Attractors $M_{MAX}$} & $M^{pp}_{MPA1}=
2.788\,M_{\odot}$ & $M^{p}_{MPA1}= 2.79107\, M_{\odot}$
\\  \hline
\textbf{Higgs Model $M_{MAX}$} & $M^{pp}_{MPA1}= 2.771\,M_{\odot}$
& $M^{p}_{MPA1}= 2.7735\, M_{\odot}$
\\  \hline
\textbf{$\alpha$-$10^4$ Attractors $M_{MAX}$} & $M^{pp}_{MPA1}=
2.749\,M_{\odot}$ & $M^{p}_{MPA1}= 2.74598\, M_{\odot}$
\\  \hline
\textbf{$\alpha$-$0.6$  Attractors $M_{MAX}$} & $M^{pp}_{MPA1}=
2.784\,M_{\odot}$ & $M^{p}_{MPA1}= 2.78707\, M_{\odot}$
\\  \hline
\textbf{Quadratic Attractors $M_{MAX}$} & $M^{pp}_{MPA1}=
2.749\,M_{\odot}$ & $M^{p}_{MPA1}= 2.75175\, M_{\odot}$
\\  \hline
    \end{tabular}
  \end{center}
\end{table}

\section*{Concluding Remarks}

In this work we studied static NSs phenomenology in the context of
various inflationary attractors using a large sample of EoSs
adopting the piecewise polytropic EoS approach. The inflationary
attractors we considered have a high cosmological significance and
specifically we considered the universal attractors, the $R^p$
attractors (three distinct model values), the induced inflation,
the quadratic inflation, the Higgs inflation and the
$a$-attractors (two distinct model values). Regarding the EoSs, we
used the WFF1, the SLy, the APR, the MS1, the AP3, the AP4, the
ENG, the MPA1 and the MS1b ones, each of which has its own
phenomenological significance. After numerically solving the TOV
equations, we extracted the Jordan frame masses $M$ and the Jordan
frame radii $R$ for all the aforementioned models and EoSs and we
constructed the corresponding $M-R$ diagrams. We also confronted
the models and EoSs with the observational and theoretical
constraints. We considered the NICER constraints and also a
modified version of the NICER constraints, which we called NICER
II, introduced in \cite{Ecker:2022dlg} which was constructed by
taking into account the heavy black-widow binary pulsar PSR
J0952-0607 with mass $M=2.35\pm 0.17$ \cite{Romani:2022jhd}. The
NICER II constraint, indicates that the radius of a
$M=1.4M_{\odot}$ must be $R_{1.4M_{\odot}}=12.33-13.25\,$km. We
also considered several theoretical constraints which are also
based on observations, which we called CSI, CSII and CSIII. The
CSI was introduced in Ref. \cite{Altiparmak:2022bke} and indicates
that the radius of an $1.4M_{\odot}$ mass NS must be
$R_{1.4M_{\odot}}=12.42^{+0.52}_{-0.99}$ while the radius of an
$2M_{\odot}$ mass NS must be
$R_{2M_{\odot}}=12.11^{+1.11}_{-1.23}\,$km. The second constraint
we considered we named it CSII and was introduced in Ref.
\cite{Raaijmakers:2021uju} and indicates that the radius of an
$1.4M_{\odot}$ mass NS must be
$R_{1.4M_{\odot}}=12.33^{+0.76}_{-0.81}\,\mathrm{km}$. Moreover,
we considered a third constraint, namely CSIII, which was
introduced in Ref. \cite{Bauswein:2017vtn} and indicates that the
radius of an $1.6M_{\odot}$ mass NS must be larger than
$R_{1.6M_{\odot}}>10.68^{+0.15}_{-0.04}\,$km, while when the
radius of a NS with maximum mass is considered, it must be larger
than $R_{M_{max}}>9.6^{+0.14}_{-0.03}\,$km.

Our results are deemed interesting and are listed below:

\begin{itemize}

    \item In the context of our work, it is possible to discriminate inflationary attractors which at the
    cosmological level are indistinguishable using the $M-R$ graphs. This feature though
    is model dependent and also EoS dependent, since for some EoSs
    and some cosmological models, the $M-R$ graphs we produced are
    identical. This for example occurred for the APR EoS and for
    the quadratic attractors, the $a$-attractors and the $R^p$-8
    attractors. Also the differences may be within the errors of
    the experimental limits, thus the discrimination has
    limitations.

    \item Among all the EoSs, the only EoS which is compatible
    with all the constraints, theoretical and observational, is
    the MPA1, for all the inflationary models considered in this work. It is remarkable that the maximum masses for this
    EoS are inside the mass-gap region, with $M>2.5M_{\odot}$, but lower than the 3 solar masses causal
    limit.

    \item As the NICER constraints are pushed towards larger
    radii, it seems that EoSs which produce maximum masses in the
    mass gap region, with $M>2.5M_{\odot}$, but lower than the 3 solar masses causal limit, are favored and
    compatible with the modified NICER constraints.

\item The only EoSs which produce maximum masses beyond the causal
limit of 3 solar masses (see Eq. (\ref{causalupperbound})) are the
MS1 and MS1b, and remarkably, none of the two produces $M-R$
graphs which are compatible with the constraints. In fact, none of
the two is compatible with even a single constraint.

\item Among all EoS we considered, the MPA1 EoS seems to play an
important role, while the WFF1, MS1 and MS1b seem to be entirely
ruled out, regarding a viable description of static NSs.

\end{itemize}
An important feature to note is that, as the NICER constraints are
pushed to higher radii, by taking into account the PSR J0952-0607,
it seems that the compatibility with the theoretical and
observational constraints comes hand in hand with models and EoSs
which generated maximum radii beyond the mass-gap region with
$M>2.5M_{\odot}$ and below the 3 solar masses causal EoSs limit.
In our case, this occurs in a flawless way for the MPA1 EoS, which
is compatible with all the constraints. This feature of course can
be met in several other distinguished cases, for example for the
ENG and AP3 EoSs and for the Higgs, the $a$-attractors and the
$R^p$ attractors models, but in the case of the MPA1 EoS we found
that all the models we considered, are compatible with the all the
theoretical and observational constraints we considered in this
article. Thus the question is, does the MPA1 EoS play an important
role in NS physics, or this multi-compatibility with the
constraints of this specific EoS is accidental? Is this EoS
fundamental for the NSs nuclear matter? We cannot tell, however it
is reportable to say the least. It is notable that in the
literature there exist articles that also find nice compatibility
properties of the MPA1 EoS with the data, see for example Ref.
\cite{Soultanis:2021oia}. We anticipate future observations to see
whether heavy NSs exist in nature, heavier that the current
$2.35M_{\odot}$ solar masses limit corresponding to the PSR
J0952-0607. If NSs are found well inside the mass-gap region,
beyond the $2.5 M_{\odot}$ limit, but to our opinion below the
causal 3 solar masses limit, this will be sensational and decisive
on whether modifications of GR are needed or not. Nature will tell
us whether GR is all what is needed for the description of
astrophysical phenomena, or whether GR is a lower limit of some
effective theory active in extreme gravity astrophysical and
cosmological phenomena. For the moment GR suffices though.

\section*{Acknowledgments}

This work was supported by MINECO (Spain), project
PID2019-104397GB-I00 (S.D.O). This work by S.D.O was also
partially supported by the program Unidad de Excelencia Maria de
Maeztu CEX2020-001058-M, Spain.


\begin{thebibliography}{99}





\bibitem{TheLIGOScientific:2017qsa}
B.~P.~Abbott \textit{et al.} [LIGO Scientific and Virgo],
Phys. Rev. Lett. \textbf{119} (2017) no.16, 161101
doi:10.1103/PhysRevLett.119.161101 [arXiv:1710.05832 [gr-qc]].


\bibitem{Abbott:2020khf}
R.~Abbott \textit{et al.} [LIGO Scientific and Virgo],
Astrophys. J. Lett. \textbf{896} (2020) no.2, L44
doi:10.3847/2041-8213/ab960f [arXiv:2006.12611 [astro-ph.HE]].



\bibitem{Ezquiaga:2017ekz}
J.~M.~Ezquiaga and M.~Zumalac\'arregui,
Phys. Rev. Lett. \textbf{119} (2017) no.25, 251304
doi:10.1103/PhysRevLett.119.251304 [arXiv:1710.05901
[astro-ph.CO]].


\bibitem{Baker:2017hug}
T.~Baker, E.~Bellini, P.~G.~Ferreira, M.~Lagos, J.~Noller and
I.~Sawicki,
Phys. Rev. Lett. \textbf{119} (2017) no.25, 251301
doi:10.1103/PhysRevLett.119.251301 [arXiv:1710.06394
[astro-ph.CO]].


\bibitem{Creminelli:2017sry}
P.~Creminelli and F.~Vernizzi,
Phys. Rev. Lett. \textbf{119} (2017) no.25, 251302
doi:10.1103/PhysRevLett.119.251302 [arXiv:1710.05877
[astro-ph.CO]].


\bibitem{Sakstein:2017xjx}
J.~Sakstein and B.~Jain,
Phys. Rev. Lett. \textbf{119} (2017) no.25, 251303
doi:10.1103/PhysRevLett.119.251303 [arXiv:1710.05893
[astro-ph.CO]].


\bibitem{Odintsov:2020sqy}
S.~D.~Odintsov, V.~K.~Oikonomou and F.~P.~Fronimos,
Nucl. Phys. B \textbf{958} (2020), 115135
doi:10.1016/j.nuclphysb.2020.115135 [arXiv:2003.13724 [gr-qc]].


\bibitem{Oikonomou:2021kql}
V.~K.~Oikonomou,
Class. Quant. Grav. \textbf{38} (2021) no.19, 195025
doi:10.1088/1361-6382/ac2168 [arXiv:2108.10460 [gr-qc]].



\bibitem{Oikonomou:2022ksx}
V.~K.~Oikonomou, P.~D.~Katzanis and I.~C.~Papadimitriou,
Class. Quant. Grav. \textbf{39} (2022) no.9, 095008
doi:10.1088/1361-6382/ac5eba [arXiv:2203.09867 [gr-qc]].



\bibitem{reviews1}
 S.~Nojiri, S.~D.~Odintsov and V.~K.~Oikonomou,
  Phys.\ Rept.\  {\bf 692} (2017) 1
  [arXiv:1705.11098 [gr-qc]].

\bibitem{reviews2}


 S. Capozziello, M. De Laurentis,
   Phys.\ Rept.\  {\bf 509}, 167 (2011);\\




\bibitem{reviews3}
 V.~Faraoni and S.~Capozziello,
  Fundam.\ Theor.\ Phys.\  {\bf 170} (2010).


   \bibitem{reviews4}

S. Nojiri, S.D. Odintsov,
   Phys.\ Rept.\  {\bf 505}, 59 (2011);


\bibitem{LIGOScientific:2020zkf}
R.~Abbott \textit{et al.} [LIGO Scientific and Virgo],
Astrophys. J. Lett. \textbf{896} (2020) no.2, L44
doi:10.3847/2041-8213/ab960f [arXiv:2006.12611 [astro-ph.HE]].



\bibitem{Romani:2022jhd}
R.~W.~Romani, D.~Kandel, A.~V.~Filippenko, T.~G.~Brink and
W.~Zheng,
Astrophys. J. Lett. \textbf{934} (2022) no.2, L18
doi:10.3847/2041-8213/ac8007 [arXiv:2207.05124 [astro-ph.HE]].














\bibitem{Haensel:2007yy}
P.~Haensel, A.~Y.~Potekhin and D.~G.~Yakovlev,
Astrophys. Space Sci. Libr. \textbf{326} (2007), pp.1-619
doi:10.1007/978-0-387-47301-7




\bibitem{Friedman:2013xza}
J.~L.~Friedman and N.~Stergioulas, ``Rotating Relativistic
Stars,'' doi:10.1017/CBO9780511977596

\bibitem{Baym:2017whm}
G.~Baym, T.~Hatsuda, T.~Kojo, P.~D.~Powell, Y.~Song and
T.~Takatsuka,
Rept. Prog. Phys. \textbf{81} (2018) no.5, 056902
doi:10.1088/1361-6633/aaae14 [arXiv:1707.04966 [astro-ph.HE]].


\bibitem{Lattimer:2004pg}
J.~M.~Lattimer and M.~Prakash,
Science \textbf{304} (2004), 536-542 doi:10.1126/science.1090720
[arXiv:astro-ph/0405262 [astro-ph]].


\bibitem{Olmo:2019flu}
G.~J.~Olmo, D.~Rubiera-Garcia and A.~Wojnar,
Phys. Rept. \textbf{876} (2020), 1-75
doi:10.1016/j.physrep.2020.07.001 [arXiv:1912.05202 [gr-qc]].















\bibitem{Lattimer:2012nd}
J.~M.~Lattimer,
Ann. Rev. Nucl. Part. Sci. \textbf{62} (2012), 485-515
doi:10.1146/annurev-nucl-102711-095018 [arXiv:1305.3510
[nucl-th]].

\bibitem{Steiner:2011ft}
A.~W.~Steiner and S.~Gandolfi,
Phys. Rev. Lett. \textbf{108} (2012), 081102
doi:10.1103/PhysRevLett.108.081102 [arXiv:1110.4142 [nucl-th]].




\bibitem{Horowitz:2005zb}
C.~J.~Horowitz, M.~A.~Perez-Garcia, D.~K.~Berry and
J.~Piekarewicz,
Phys. Rev. C \textbf{72} (2005), 035801
doi:10.1103/PhysRevC.72.035801 [arXiv:nucl-th/0508044 [nucl-th]].


\bibitem{Watanabe:2000rj}
G.~Watanabe, K.~Iida and K.~Sato,
Nucl. Phys. A \textbf{676} (2000), 455-473 [erratum: Nucl. Phys. A
\textbf{726} (2003), 357-365] doi:10.1016/S0375-9474(00)00197-4
[arXiv:astro-ph/0001273 [astro-ph]].

\bibitem{Shen:1998gq}
H.~Shen, H.~Toki, K.~Oyamatsu and K.~Sumiyoshi,
Nucl. Phys. A \textbf{637} (1998), 435-450
doi:10.1016/S0375-9474(98)00236-X [arXiv:nucl-th/9805035
[nucl-th]].




\bibitem{Xu:2009vi}
J.~Xu, L.~W.~Chen, B.~A.~Li and H.~R.~Ma,
Astrophys. J. \textbf{697} (2009), 1549-1568
doi:10.1088/0004-637X/697/2/1549 [arXiv:0901.2309 [astro-ph.SR]].


\bibitem{Hebeler:2013nza}
K.~Hebeler, J.~M.~Lattimer, C.~J.~Pethick and A.~Schwenk,
Astrophys. J. \textbf{773} (2013), 11
doi:10.1088/0004-637X/773/1/11 [arXiv:1303.4662 [astro-ph.SR]].


\bibitem{Mendoza-Temis:2014mja}
J.~de Jes\'us Mendoza-Temis, M.~R.~Wu, G.~Mart\'\i{}nez-Pinedo,
K.~Langanke, A.~Bauswein and H.~T.~Janka,
Phys. Rev. C \textbf{92} (2015) no.5, 055805
doi:10.1103/PhysRevC.92.055805 [arXiv:1409.6135 [astro-ph.HE]].





\bibitem{Ho:2014pta}
W.~C.~G.~Ho, K.~G.~Elshamouty, C.~O.~Heinke and A.~Y.~Potekhin,
Phys. Rev. C \textbf{91} (2015) no.1, 015806
doi:10.1103/PhysRevC.91.015806 [arXiv:1412.7759 [astro-ph.HE]].


\bibitem{Kanakis-Pegios:2020kzp}
A.~Kanakis-Pegios, P.~S.~Koliogiannis and C.~C.~Moustakidis,
[arXiv:2012.09580 [astro-ph.HE]].


\bibitem{Tsaloukidis:2022rus}
L.~Tsaloukidis, P.~S.~Koliogiannis, A.~Kanakis-Pegios and
C.~C.~Moustakidis,
[arXiv:2210.15644 [astro-ph.HE]].




\bibitem{Buschmann:2019pfp}
M.~Buschmann, R.~T.~Co, C.~Dessert and B.~R.~Safdi,
Phys. Rev. Lett. \textbf{126} (2021) no.2, 021102
doi:10.1103/PhysRevLett.126.021102 [arXiv:1910.04164 [hep-ph]].




\bibitem{Safdi:2018oeu}
B.~R.~Safdi, Z.~Sun and A.~Y.~Chen,
Phys. Rev. D \textbf{99} (2019) no.12, 123021
doi:10.1103/PhysRevD.99.123021 [arXiv:1811.01020 [astro-ph.CO]].



\bibitem{Hook:2018iia}
A.~Hook, Y.~Kahn, B.~R.~Safdi and Z.~Sun,
Phys. Rev. Lett. \textbf{121} (2018) no.24, 241102
doi:10.1103/PhysRevLett.121.241102 [arXiv:1804.03145 [hep-ph]].


\bibitem{Edwards:2020afl}
T.~D.~P.~Edwards, B.~J.~Kavanagh, L.~Visinelli and C.~Weniger,
[arXiv:2011.05378 [hep-ph]].

\bibitem{Nurmi:2021xds}
S.~Nurmi, E.~D.~Schiappacasse and T.~T.~Yanagida,
[arXiv:2102.05680 [hep-ph]].


\bibitem{Astashenok:2020qds}
A.~V.~Astashenok, S.~Capozziello, S.~D.~Odintsov and
V.~K.~Oikonomou,
Phys. Lett. B \textbf{811} (2020), 135910
doi:10.1016/j.physletb.2020.135910 [arXiv:2008.10884 [gr-qc]].



\bibitem{Astashenok:2021peo}
A.~V.~Astashenok, S.~Capozziello, S.~D.~Odintsov and
V.~K.~Oikonomou,
[arXiv:2103.04144 [gr-qc]].

\bibitem{Capozziello:2015yza}
S.~Capozziello, M.~De Laurentis, R.~Farinelli and S.~D.~Odintsov,
Phys. Rev. D \textbf{93} (2016) no.2, 023501
doi:10.1103/PhysRevD.93.023501 [arXiv:1509.04163 [gr-qc]].


\bibitem{Astashenok:2014nua}
A.~V.~Astashenok, S.~Capozziello and S.~D.~Odintsov,
JCAP \textbf{01} (2015), 001 doi:10.1088/1475-7516/2015/01/001
[arXiv:1408.3856 [gr-qc]].


\bibitem{Astashenok:2014pua}
A.~V.~Astashenok, S.~Capozziello and S.~D.~Odintsov,
Phys. Rev. D \textbf{89} (2014) no.10, 103509
doi:10.1103/PhysRevD.89.103509 [arXiv:1401.4546 [gr-qc]].



\bibitem{Astashenok:2013vza}
A.~V.~Astashenok, S.~Capozziello and S.~D.~Odintsov,
JCAP \textbf{12} (2013), 040 doi:10.1088/1475-7516/2013/12/040
[arXiv:1309.1978 [gr-qc]].




\bibitem{Arapoglu:2010rz}
A.~S.~Arapoglu, C.~Deliduman and K.~Y.~Eksi,
JCAP \textbf{07} (2011), 020 doi:10.1088/1475-7516/2011/07/020
[arXiv:1003.3179 [gr-qc]].


\bibitem{Panotopoulos:2021sbf}
G.~Panotopoulos, T.~Tangphati, A.~Banerjee and M.~K.~Jasim,
[arXiv:2104.00590 [gr-qc]].


\bibitem{Lobato:2020fxt}
R.~Lobato, O.~Louren\c{c}o, P.~H.~R.~S.~Moraes, C.~H.~Lenzi, M.~de
Avellar, W.~de Paula, M.~Dutra and M.~Malheiro,
JCAP \textbf{12} (2020), 039 doi:10.1088/1475-7516/2020/12/039
[arXiv:2009.04696 [astro-ph.HE]].


\bibitem{Numajiri:2021nsc}
K.~Numajiri, T.~Katsuragawa and S.~Nojiri,
Phys. Lett. B \textbf{826} (2022), 136929
doi:10.1016/j.physletb.2022.136929 [arXiv:2111.02660 [gr-qc]].




\bibitem{Altiparmak:2022bke}
S.~Altiparmak, C.~Ecker and L.~Rezzolla,
[arXiv:2203.14974 [astro-ph.HE]].

\bibitem{Bauswein:2020kor}
A.~Bauswein, G.~Guo, J.~H.~Lien, Y.~H.~Lin and M.~R.~Wu,
[arXiv:2012.11908 [astro-ph.HE]].


\bibitem{Vretinaris:2019spn}
S.~Vretinaris, N.~Stergioulas and A.~Bauswein,
Phys. Rev. D \textbf{101} (2020) no.8, 084039
doi:10.1103/PhysRevD.101.084039 [arXiv:1910.10856 [gr-qc]].



\bibitem{Bauswein:2020aag}
A.~Bauswein, S.~Blacker, V.~Vijayan, N.~Stergioulas,
K.~Chatziioannou, J.~A.~Clark, N.~U.~F.~Bastian, D.~B.~Blaschke,
M.~Cierniak and T.~Fischer,
Phys. Rev. Lett. \textbf{125} (2020) no.14, 141103
doi:10.1103/PhysRevLett.125.141103 [arXiv:2004.00846
[astro-ph.HE]].


\bibitem{Bauswein:2017vtn}
A.~Bauswein, O.~Just, H.~T.~Janka and N.~Stergioulas,
Astrophys. J. Lett. \textbf{850} (2017) no.2, L34
doi:10.3847/2041-8213/aa9994 [arXiv:1710.06843 [astro-ph.HE]].



\bibitem{Most:2018hfd}
E.~R.~Most, L.~R.~Weih, L.~Rezzolla and J.~Schaffner-Bielich,
Phys. Rev. Lett. \textbf{120} (2018) no.26, 261103
doi:10.1103/PhysRevLett.120.261103 [arXiv:1803.00549 [gr-qc]].



\bibitem{Rezzolla:2017aly}
L.~Rezzolla, E.~R.~Most and L.~R.~Weih,
Astrophys. J. Lett. \textbf{852} (2018) no.2, L25
doi:10.3847/2041-8213/aaa401 [arXiv:1711.00314 [astro-ph.HE]].



\bibitem{Nathanail:2021tay}
A.~Nathanail, E.~R.~Most and L.~Rezzolla,
Astrophys. J. Lett. \textbf{908} (2021) no.2, L28
doi:10.3847/2041-8213/abdfc6 [arXiv:2101.01735 [astro-ph.HE]].


\bibitem{Koppel:2019pys}
S.~K\"oppel, L.~Bovard and L.~Rezzolla,
Astrophys. J. Lett. \textbf{872} (2019) no.1, L16
doi:10.3847/2041-8213/ab0210 [arXiv:1901.09977 [gr-qc]].


\bibitem{Raaijmakers:2021uju}
G.~Raaijmakers, S.~K.~Greif, K.~Hebeler, T.~Hinderer, S.~Nissanke,
A.~Schwenk, T.~E.~Riley, A.~L.~Watts, J.~M.~Lattimer and
W.~C.~G.~Ho,
Astrophys. J. Lett. \textbf{918} (2021) no.2, L29
doi:10.3847/2041-8213/ac089a [arXiv:2105.06981 [astro-ph.HE]].


\bibitem{Most:2020exl}
E.~R.~Most, L.~J.~Papenfort, S.~Tootle and L.~Rezzolla,
Astrophys. J. \textbf{912} (2021) no.1, 80
doi:10.3847/1538-4357/abf0a5 [arXiv:2012.03896 [astro-ph.HE]].

\bibitem{Ecker:2022dlg}
C.~Ecker and L.~Rezzolla,
[arXiv:2209.08101 [astro-ph.HE]].


\bibitem{Jiang:2022tps}
J.~L.~Jiang, C.~Ecker and L.~Rezzolla,
[arXiv:2211.00018 [gr-qc]].











\bibitem{Pani:2014jra}
P.~Pani and E.~Berti,
Phys. Rev. D \textbf{90} (2014) no.2, 024025
doi:10.1103/PhysRevD.90.024025 [arXiv:1405.4547 [gr-qc]].


\bibitem{Staykov:2014mwa}
K.~V.~Staykov, D.~D.~Doneva, S.~S.~Yazadjiev and K.~D.~Kokkotas,
JCAP \textbf{10} (2014), 006 doi:10.1088/1475-7516/2014/10/006
[arXiv:1407.2180 [gr-qc]].





\bibitem{Horbatsch:2015bua}
M.~Horbatsch, H.~O.~Silva, D.~Gerosa, P.~Pani, E.~Berti,
L.~Gualtieri and U.~Sperhake,
Class. Quant. Grav. \textbf{32} (2015) no.20, 204001
doi:10.1088/0264-9381/32/20/204001 [arXiv:1505.07462 [gr-qc]].



\bibitem{Silva:2014fca}
H.~O.~Silva, C.~F.~B.~Macedo, E.~Berti and L.~C.~B.~Crispino,
Class. Quant. Grav. \textbf{32} (2015), 145008
doi:10.1088/0264-9381/32/14/145008 [arXiv:1411.6286 [gr-qc]].

\bibitem{Doneva:2013qva}
D.~D.~Doneva, S.~S.~Yazadjiev, N.~Stergioulas and K.~D.~Kokkotas,
Phys. Rev. D \textbf{88} (2013) no.8, 084060
doi:10.1103/PhysRevD.88.084060 [arXiv:1309.0605 [gr-qc]].




\bibitem{Xu:2020vbs}
R.~Xu, Y.~Gao and L.~Shao,
Phys. Rev. D \textbf{102} (2020) no.6, 064057
doi:10.1103/PhysRevD.102.064057 [arXiv:2007.10080 [gr-qc]].


\bibitem{Salgado:1998sg}
M.~Salgado, D.~Sudarsky and U.~Nucamendi,
Phys. Rev. D \textbf{58} (1998), 124003
doi:10.1103/PhysRevD.58.124003 [arXiv:gr-qc/9806070 [gr-qc]].



\bibitem{Shibata:2013pra}
M.~Shibata, K.~Taniguchi, H.~Okawa and A.~Buonanno,
Phys. Rev. D \textbf{89} (2014) no.8, 084005
doi:10.1103/PhysRevD.89.084005 [arXiv:1310.0627 [gr-qc]].



\bibitem{Arapoglu:2019mun}
A.~Sava\c{s} Arapo\u{g}lu, K.~Yavuz Ek\c{s}i and A.~Emrah
Y\"ukselci,
Phys. Rev. D \textbf{99} (2019) no.6, 064055
doi:10.1103/PhysRevD.99.064055 [arXiv:1903.00391 [gr-qc]].

\bibitem{Ramazanoglu:2016kul}
F.~M.~Ramazano\u{g}lu and F.~Pretorius,
Phys. Rev. D \textbf{93} (2016) no.6, 064005
doi:10.1103/PhysRevD.93.064005 [arXiv:1601.07475 [gr-qc]].

\bibitem{AltahaMotahar:2019ekm}
Z.~Altaha Motahar, J.~L.~Bl\'azquez-Salcedo, D.~D.~Doneva, J.~Kunz
and S.~S.~Yazadjiev,
Phys. Rev. D \textbf{99} (2019) no.10, 104006
doi:10.1103/PhysRevD.99.104006 [arXiv:1902.01277 [gr-qc]].



\bibitem{Chew:2019lsa}
X.~Y.~Chew, V.~Dzhunushaliev, V.~Folomeev, B.~Kleihaus and
J.~Kunz,
Phys. Rev. D \textbf{100} (2019) no.4, 044019
doi:10.1103/PhysRevD.100.044019 [arXiv:1906.08742 [gr-qc]].



\bibitem{Blazquez-Salcedo:2020ibb}
J.~L.~Bl\'azquez-Salcedo, F.~Scen Khoo and J.~Kunz,
EPL \textbf{130} (2020) no.5, 50002
doi:10.1209/0295-5075/130/50002 [arXiv:2001.09117 [gr-qc]].





\bibitem{Motahar:2017blm}
Z.~Altaha Motahar, J.~L.~Bl\'azquez-Salcedo, B.~Kleihaus and
J.~Kunz,
Phys. Rev. D \textbf{96} (2017) no.6, 064046
doi:10.1103/PhysRevD.96.064046 [arXiv:1707.05280 [gr-qc]].



\bibitem{Odintsov:2021qbq}
S.~D.~Odintsov and V.~K.~Oikonomou,
Phys. Dark Univ. \textbf{32} (2021), 100805
doi:10.1016/j.dark.2021.100805 [arXiv:2103.07725 [gr-qc]].



\bibitem{Odintsov:2021nqa}
S.~D.~Odintsov and V.~K.~Oikonomou,
Annals Phys. \textbf{440} (2022), 168839
doi:10.1016/j.aop.2022.168839 [arXiv:2104.01982 [gr-qc]].



\bibitem{Oikonomou:2021iid}
V.~K.~Oikonomou,
Class. Quant. Grav. \textbf{38} (2021) no.17, 175005
doi:10.1088/1361-6382/ac161c [arXiv:2107.12430 [gr-qc]].

\bibitem{Pretel:2022rwx}
J.~M.~Z.~Pretel, J.~D.~V.~Arba\~nil, S.~B.~Duarte, S.~E.~Jor\'as
and R.~R.~R.~Reis,
[arXiv:2206.03878 [gr-qc]].

\bibitem{Pretel:2022plg}
J.~M.~Z.~Pretel and S.~B.~Duarte,
[arXiv:2202.04467 [gr-qc]].


\bibitem{Cuzinatto:2016ehv}
R.~R.~Cuzinatto, C.~A.~M.~de Melo, L.~G.~Medeiros and
P.~J.~Pompeia,
Phys. Rev. D \textbf{93} (2016) no.12, 124034 [erratum: Phys. Rev.
D \textbf{98} (2018) no.2, 029901] doi:10.1103/PhysRevD.93.124034
[arXiv:1603.01563 [gr-qc]].



\bibitem{Oikonomou:2023dgu}
V.~K.~Oikonomou,
[arXiv:2301.12136 [gr-qc]].



\bibitem{Akrami:2018odb}
Y.~Akrami \textit{et al.} [Planck],
Astron. Astrophys. \textbf{641} (2020), A10
doi:10.1051/0004-6361/201833887 [arXiv:1807.06211 [astro-ph.CO]].











\bibitem{alpha0}
R.~Kallosh, A.~Linde and D.~Roest,
JHEP \textbf{09} (2014), 062 doi:10.1007/JHEP09(2014)062
[arXiv:1407.4471 [hep-th]].



\bibitem{alpha1} R.~Kallosh and A.~Linde,
  JCAP {\bf 1307} (2013) 002
  [arXiv:1306.5220 [hep-th]].




\bibitem{alpha2} S.~Ferrara, R.~Kallosh, A.~Linde and M.~Porrati,
  Phys.\ Rev.\ D {\bf 88} (2013) no.8,  085038
  [arXiv:1307.7696 [hep-th]].


\bibitem{alpha3}R.~Kallosh, A.~Linde and D.~Roest,
  JHEP {\bf 1311} (2013) 198
  [arXiv:1311.0472 [hep-th]].






\bibitem{alpha4}

JCAP \textbf{05} (2015), 003 doi:10.1088/1475-7516/2015/05/003
[arXiv:1504.00663 [hep-th]].






\bibitem{alpha5} S.~Cecotti and R.~Kallosh,
  JHEP {\bf 1405} (2014) 114
  [arXiv:1403.2932 [hep-th]].



\bibitem{alpha6} J.~J.~M.~Carrasco, R.~Kallosh and A.~Linde,
  JHEP {\bf 1510} (2015) 147
  [arXiv:1506.01708 [hep-th]].



\bibitem{alpha7}   J.~J.~M.~Carrasco, R.~Kallosh, A.~Linde and D.~Roest,
Phys. Rev. D \textbf{92} (2015) no.4, 041301
doi:10.1103/PhysRevD.92.041301 [arXiv:1504.05557 [hep-th]].


\bibitem{alpha7a}

R.~Kallosh, A.~Linde and D.~Roest,
Phys. Rev. Lett. \textbf{112} (2014) no.1, 011303
doi:10.1103/PhysRevLett.112.011303 [arXiv:1310.3950 [hep-th]].


\bibitem{alpha8} D.~Roest and M.~Scalisi,
  Phys.\ Rev.\ D {\bf 92} (2015) 043525
  doi:10.1103/PhysRevD.92.043525
  [arXiv:1503.07909 [hep-th]].




\bibitem{alpha9}  R.~Kallosh, A.~Linde and D.~Roest,
  JHEP {\bf 1408} (2014) 052
  doi:10.1007/JHEP08(2014)052
  [arXiv:1405.3646 [hep-th]].

\bibitem{alpha10} J.~Ellis, D.~V.~Nanopoulos and K.~A.~Olive,
  JCAP {\bf 1310} (2013) 009
  [arXiv:1307.3537 [hep-th]].

\bibitem{alpha11} Y.~F.~Cai, J.~O.~Gong and S.~Pi,
  Phys.\ Lett.\ B {\bf 738} (2014) 20
  doi:10.1016/j.physletb.2014.09.009
  [arXiv:1404.2560 [hep-th]].

\bibitem{alpha12}
  Z.~Yi and Y.~Gong,
  arXiv:1608.05922 [gr-qc].


\bibitem{alpha13}

Y.~Akrami, R.~Kallosh, A.~Linde and V.~Vardanyan,
JCAP \textbf{06} (2018), 041 doi:10.1088/1475-7516/2018/06/041
[arXiv:1712.09693 [hep-th]].



\bibitem{alpha14}
S.~Qummer, A.~Jawad and M.~Younas,
Int. J. Mod. Phys. D \textbf{29} (2020) no.16, 2050117
doi:10.1142/S0218271820501175



\bibitem{alpha15}
Q.~Fei, Z.~Yi and Y.~Yang,
Universe \textbf{6} (2020) no.11, 213 doi:10.3390/universe6110213
[arXiv:2009.14819 [gr-qc]].



\bibitem{alpha16}
A.~D.~Kanfon, F.~Mavoa and S.~M.~J.~Houndjo,
Astrophys. Space Sci. \textbf{365} (2020) no.6, 97
doi:10.1007/s10509-020-03813-6


\bibitem{alpha17}
I.~Antoniadis, A.~Karam, A.~Lykkas, T.~Pappas and K.~Tamvakis,
PoS \textbf{CORFU2019} (2020), 073 doi:10.22323/1.376.0073
[arXiv:1912.12757 [gr-qc]].


\bibitem{alpha18}
C.~Garc\'\i{}a-Garc\'\i{}a, P.~Ru\'\i{}z-Lapuente, D.~Alonso and
M.~Zumalac\'arregui,
JCAP \textbf{07} (2019), 025 doi:10.1088/1475-7516/2019/07/025
[arXiv:1905.03753 [astro-ph.CO]].



\bibitem{alpha19}
F.~X.~Linares Cede\~no, A.~Montiel, J.~C.~Hidalgo and G.~Germ\'an,
JCAP \textbf{08} (2019), 002 doi:10.1088/1475-7516/2019/08/002
[arXiv:1905.00834 [gr-qc]].



\bibitem{alpha20}
S.~Karamitsos,
JCAP \textbf{09} (2019), 022 doi:10.1088/1475-7516/2019/09/022
[arXiv:1903.03707 [hep-th]].


\bibitem{alpha21}
D.~D.~Canko, I.~D.~Gialamas and G.~P.~Kodaxis,
Eur. Phys. J. C \textbf{80} (2020) no.5, 458
doi:10.1140/epjc/s10052-020-8025-4 [arXiv:1901.06296 [hep-th]].


\bibitem{alpha22}
T.~Miranda, C.~Escamilla-Rivera, O.~F.~Piattella and J.~C.~Fabris,
JCAP \textbf{05} (2019), 028 doi:10.1088/1475-7516/2019/05/028
[arXiv:1812.01287 [gr-qc]].


\bibitem{alpha23}
A.~Karam, T.~Pappas and K.~Tamvakis,
JCAP \textbf{02} (2019), 006 doi:10.1088/1475-7516/2019/02/006
[arXiv:1810.12884 [gr-qc]].


\bibitem{alpha24}
K.~Nozari and N.~Rashidi,
Astrophys. J. \textbf{863} (2018) no.2, 133
doi:10.3847/1538-4357/aad18e [arXiv:1808.05363 [astro-ph.CO]].


\bibitem{alpha25}
C.~Garc\'\i{}a-Garc\'\i{}a, E.~V.~Linder, P.~Ru\'\i{}z-Lapuente
and M.~Zumalac\'arregui,
JCAP \textbf{08} (2018), 022 doi:10.1088/1475-7516/2018/08/022
[arXiv:1803.00661 [astro-ph.CO]].



\bibitem{alpha26}
N.~Rashidi and K.~Nozari,
Int. J. Mod. Phys. D \textbf{27} (2018) no.07, 1850076
doi:10.1142/S0218271818500761 [arXiv:1802.09185 [astro-ph.CO]].



\bibitem{alpha27}
Q.~Gao, Y.~Gong and Q.~Fei,
JCAP \textbf{05} (2018), 005 doi:10.1088/1475-7516/2018/05/005
[arXiv:1801.09208 [gr-qc]].


\bibitem{alpha28}
K.~Dimopoulos, L.~Donaldson Wood and C.~Owen,
Phys. Rev. D \textbf{97} (2018) no.6, 063525
doi:10.1103/PhysRevD.97.063525 [arXiv:1712.01760 [astro-ph.CO]].



\bibitem{alpha29}
T.~Miranda, J.~C.~Fabris and O.~F.~Piattella,
JCAP \textbf{09} (2017), 041 doi:10.1088/1475-7516/2017/09/041
[arXiv:1707.06457 [gr-qc]].



\bibitem{alpha30}
A.~Karam, T.~Pappas and K.~Tamvakis,
Phys. Rev. D \textbf{96} (2017) no.6, 064036
doi:10.1103/PhysRevD.96.064036 [arXiv:1707.00984 [gr-qc]].



\bibitem{alpha31}
K.~Nozari and N.~Rashidi,
Phys. Rev. D \textbf{95} (2017) no.12, 123518
doi:10.1103/PhysRevD.95.123518 [arXiv:1705.02617 [astro-ph.CO]].



\bibitem{alpha32}
Q.~Gao and Y.~Gong,
Eur. Phys. J. Plus \textbf{133} (2018) no.11, 491
doi:10.1140/epjp/i2018-12324-3 [arXiv:1703.02220 [gr-qc]].


\bibitem{alpha33}
C.~Q.~Geng, C.~C.~Lee and Y.~P.~Wu,
Eur. Phys. J. C \textbf{77} (2017) no.3, 162
doi:10.1140/epjc/s10052-017-4720-1 [arXiv:1512.04019
[astro-ph.CO]].


\bibitem{alpha34}
S.~D.~Odintsov and V.~K.~Oikonomou,
Phys. Lett. B \textbf{807} (2020), 135576
doi:10.1016/j.physletb.2020.135576 [arXiv:2005.12804 [gr-qc]].



\bibitem{alpha35}
S.~D.~Odintsov and V.~K.~Oikonomou,
Phys. Rev. D \textbf{94} (2016) no.12, 124026
doi:10.1103/PhysRevD.94.124026 [arXiv:1612.01126 [gr-qc]].



\bibitem{alpha36}
S.~D.~Odintsov and V.~K.~Oikonomou,
Class. Quant. Grav. \textbf{34} (2017) no.10, 105009
doi:10.1088/1361-6382/aa69a8 [arXiv:1611.00738 [gr-qc]].


\bibitem{alpha37}

L.~J\"arv, A.~Karam, A.~Kozak, A.~Lykkas, A.~Racioppi and M.~Saal,
Phys. Rev. D \textbf{102} (2020) no.4, 044029
doi:10.1103/PhysRevD.102.044029 [arXiv:2005.14571 [gr-qc]].









\bibitem{Read:2008iy}
J.~S.~Read, B.~D.~Lackey, B.~J.~Owen and J.~L.~Friedman,
Phys. Rev. D \textbf{79} (2009), 124032


\bibitem{Read:2009yp}
J.~S.~Read, C.~Markakis, M.~Shibata, K.~Uryu, J.~D.~E.~Creighton
and J.~L.~Friedman,
Phys. Rev. D \textbf{79} (2009), 124033



\bibitem{Douchin:2001sv}
F.~Douchin and P.~Haensel,
Astron. Astrophys. \textbf{380} (2001), 151
doi:10.1051/0004-6361:20011402 [arXiv:astro-ph/0111092
[astro-ph]].



\bibitem{Akmal:1998cf}
A.~Akmal, V.~R.~Pandharipande and D.~G.~Ravenhall,
Phys. Rev. C \textbf{58} (1998), 1804-1828
doi:10.1103/PhysRevC.58.1804 [arXiv:nucl-th/9804027 [nucl-th]].

\bibitem{Wiringa:1988tp}
R.~B.~Wiringa, V.~Fiks and A.~Fabrocini,
Phys. Rev. C \textbf{38} (1988), 1010-1037
doi:10.1103/PhysRevC.38.1010




\bibitem{Engvik:1995gn}
L.~Engvik, G.~Bao, M.~Hjorth-Jensen, E.~Osnes and E.~Ostgaard,
Astrophys. J. \textbf{469} (1996), 794 doi:10.1086/177827
[arXiv:nucl-th/9509016 [nucl-th]].


\bibitem{Muther:1987xaa}
H.~M\"uther, M.~Prakash and T.~L.~Ainsworth,
Phys. Lett. B \textbf{199} (1987), 469-474
doi:10.1016/0370-2693(87)91611-X


\bibitem{Mueller:1996pm}
H.~Mueller and B.~D.~Serot,
Nucl. Phys. A \textbf{606} (1996), 508-537
doi:10.1016/0375-9474(96)00187-X [arXiv:nucl-th/9603037
[nucl-th]].


\bibitem{Akmal:1997ft}
A.~Akmal and V.~R.~Pandharipande,
Phys. Rev. C \textbf{56} (1997), 2261-2279
doi:10.1103/PhysRevC.56.2261 [arXiv:nucl-th/9705013 [nucl-th]].








\bibitem{Arnowitt:1960zzc}
R.~Arnowitt, S.~Deser and C.~W.~Misner,
Phys. Rev. \textbf{118} (1960), 1100-1104
doi:10.1103/PhysRev.118.1100



\bibitem{Miller:2021qha}
M.~C.~Miller, F.~K.~Lamb, A.~J.~Dittmann, S.~Bogdanov,
Z.~Arzoumanian, K.~C.~Gendreau, S.~Guillot, W.~C.~G.~Ho,
J.~M.~Lattimer and M.~Loewenstein, \textit{et al.}
Astrophys. J. Lett. \textbf{918} (2021) no.2, L28
doi:10.3847/2041-8213/ac089b [arXiv:2105.06979 [astro-ph.HE]].

\bibitem{Kaiser:1994vs}
D.~I.~Kaiser,
Phys. Rev. D \textbf{52} (1995), 4295-4306
doi:10.1103/PhysRevD.52.4295 [arXiv:astro-ph/9408044 [astro-ph]].

\bibitem{valerio} Valerio Faraoni, Cosmology in Scalar-Tensor Gravity,
Springer 2004

\bibitem{Faraoni:2013igs}
V.~Faraoni,
Galaxies \textbf{1} (2013) no.2, 96-106
doi:10.3390/galaxies1020096

\bibitem{Buck:2010sv}
M.~Buck, M.~Fairbairn and M.~Sakellariadou,
Phys. Rev. D \textbf{82} (2010), 043509
doi:10.1103/PhysRevD.82.043509 [arXiv:1005.1188 [hep-th]].


\bibitem{ref1}
V.~Faraoni,
Phys. Rev. D \textbf{53} (1996), 6813-6821
doi:10.1103/PhysRevD.53.6813 [arXiv:astro-ph/9602111 [astro-ph]].


\bibitem{ref2}
S.~Sonego and V.~Faraoni,
Class. Quant. Grav. \textbf{10} (1993), 1185-1187
doi:10.1088/0264-9381/10/6/015


\bibitem{OBoyle:2020qvf}
M.~F.~O'Boyle, C.~Markakis, N.~Stergioulas and J.~S.~Read,
Phys. Rev. D \textbf{102} (2020) no.8, 083027
doi:10.1103/PhysRevD.102.083027 [arXiv:2008.03342 [astro-ph.HE]].


\bibitem{Lindblom:2010bb}
L.~Lindblom,
Phys. Rev. D \textbf{82} (2010), 103011
doi:10.1103/PhysRevD.82.103011 [arXiv:1009.0738 [astro-ph.HE]].


\bibitem{submitted} Neutron Stars in Induced inflation and
Quadratic Inflation Attractors, V.K. Oikonomou, submitted for
publication.




\bibitem{Odintsov:2022bpg}
S.~D.~Odintsov and V.~K.~Oikonomou,
[arXiv:2210.11351 [gr-qc]].


\bibitem{Mishra:2018dtg}
S.~S.~Mishra, V.~Sahni and A.~V.~Toporensky,
Phys. Rev. D \textbf{98} (2018) no.8, 083538
doi:10.1103/PhysRevD.98.083538 [arXiv:1801.04948 [gr-qc]].





\bibitem{niksterg} Nikolaos Stergioulas, https://github.com/niksterg



\bibitem{Rhoades:1974fn}
C.~E.~Rhoades, Jr. and R.~Ruffini,
Phys. Rev. Lett. \textbf{32} (1974), 324-327
doi:10.1103/PhysRevLett.32.324


\bibitem{Kalogera:1996ci}
V.~Kalogera and G.~Baym,
Astrophys. J. Lett. \textbf{470} (1996), L61-L64
doi:10.1086/310296 [arXiv:astro-ph/9608059 [astro-ph]].


\bibitem{Soultanis:2021oia}
T.~Soultanis, A.~Bauswein and N.~Stergioulas,
Phys. Rev. D \textbf{105} (2022) no.4, 043020
doi:10.1103/PhysRevD.105.043020 [arXiv:2111.08353 [astro-ph.HE]].

\end{thebibliography}
\end{document}